\documentclass[sn-mathphys]{sn-jnl}

\jyear{2021}%

\theoremstyle{thmstyleone}%
\theoremstyle{thmstyletwo}%

\theoremstyle{thmstylethree}%
\usepackage{verbatim}
\usepackage{adjustbox}
\usepackage{multicol}
\usepackage{subfig}

\begin{document}

\title[Optimal dynamic climate adaptation pathways]{Optimal dynamic climate adaptation pathways: a case study of New York City}


\author*[1]{\fnm{Chi} \sur{Truong}}\email{chi.truong@mq.edu.au}

\author[2]{\fnm{Matteo} \sur{Malavasi}}\email{m.malavasi@unsw.edu.au}

\author[3]{\fnm{Han} \sur{Li}}\email{han.li@unimelb.edu.au}

\author*[1]{\fnm{Stefan} \sur{Tr{\"u}ck}}\email{stefan.trueck@mq.edu.au}

\author[1]{\fnm{Pavel V.} \sur{Shevchenko}}\email{pavel.shevchenko@mq.edu.au}

\affil[1]{\orgdiv{\normalsize Department of Actuarial Studies and Business Analytics},
\orgname{Macquarie Business School, Macquarie University}, \orgaddress{\state{NSW}, \country{Australia}}}

\affil[2]{\orgdiv{\normalsize School of Risk and Actuarial Studies}, \orgname{\normalsize UNSW Sydney}, \orgaddress{\state{NSW}, \country{\normalsize Australia}}}

\affil[3]{\orgdiv{\normalsize Department of Economics}, \orgname{\normalsize University of Melbourne}, \orgaddress{\state{VIC}, \country{\normalsize Australia}}}


\abstract{\textcolor{black}{Assessing climate risk and its potential impacts on our cities and economies is of fundamental importance. Extreme weather events, such as hurricanes, floods, and storm surges can lead to catastrophic damages. We propose a flexible approach based on real options analysis and extreme value theory, which enables the selection of optimal adaptation pathways for a portfolio of climate adaptation projects. We model the severity of extreme sea level events using the block maxima approach from extreme value theory, and then develop a real options framework, factoring in climate change, sea level rise uncertainty, and the growth in asset exposure. We then apply the proposed framework to a real-world problem, considering sea level data as well as different adaptation investment options for New York City. Our research can assist governments and policy makers in taking informed decisions about optimal adaptation pathways and more specifically about reducing flood and storm surge risk in a dynamic settings.}}




\maketitle
\clearpage
\newpage

\section{Introduction}\label{sec1}

Catastrophes such as hurricanes, floods, and storm surges have caused significant damages in coastal regions. In the presence of population growth and increasing value of physical assets in coastal areas, the frequency and severity of losses from catastrophic events can be expected to increase further with the impact of climate change \cite{portner2022climate}. For example, in 2005 in the US, Hurricane Katrina claimed more than 1800 lives and caused a total damage of \$108 billion \citep{wang2015adaptation}, while the total costs of recent hurricanes Harvey and Irma also exceeded \$100 billion. These catastrophes have increased the debt under the National Flood Insurance Program (NFIP) in the US to nearly \$20 billion and have raised serious concerns about the sustainability of the program \citep{rade2017}.  Optimal planning for adaptation to these emerging risks is therefore paramount for coastal regions \cite[see, for example][]{song2019,al2020}. It is particularly important for those regions with considerable wealth and concentration of exposure such as New York City (NYC), New Orleans, and South Florida.

Climate adaptation projects for reducing flood and storm surge risk in coastal areas include hard protection measures and soft measures. Hard protection measures such as dikes or sea walls aim to protect the region up to the designed flood height. Soft measures include building codes that require buildings to be elevated to a certain height, dry flood-proofing measures that seal the buildings to prevent water from entering into the buildings, and wet flood-proofing measures that move valuable assets to a higher level. Soft measures also include land use policies that convert the most vulnerable housing areas to park land. Among these measures, decision makers are often asked to choose from several alternative and/or complementary projects. Under these premises, flexible evaluation tools should be applied to provide useful insights to decision makers.

The value of flexibility in decision making under climate change uncertainty has been recognized in so-called 'adaptation pathways' studies \cite[see,][]{fazey2016,buurman2016,werners2021,cradock2020}. Adaptation pathways provide insights into the sequencing of actions over time, aiming to construct flexible strategies that can then be adapted as new information is revealed, or when the underlying conditions have changed \cite[see, e.g.,][]{haasnoot2013,ranger2013,kim2019}. To deal with uncertainty about the future, the possibility to spread, expand, or even abandon an investment over time should be considered \citep{dobes2010,haasnoot2013,nicholls2014}. Adaptation pathways have been successfully applied to complex problems in a wide range of fields, including biodiversity, natural resource management, and coastal development \cite[see, among others,][]{downing2012,haasnoot2012,haasnoot2013,wise2014, fazey2016}. Despite its great success, the adaptation pathways framework has typically been applied in a qualitative way only, where the selected adaptation actions can be strongly subjective.

In contrast to a more qualitative adaptation pathways approach, real option analysis is a quantitative tool that can be used to determine the value of flexibility under climate change uncertainty. Real option analysis is a popular approach to climate adaptation studies, and has been implemented in various settings \cite[see, ][]{brandao2005,dobes2010,woodward2011,woodward2014,wreford2020,park2014,chan2016dynamic,oh2018investment,park2014,kim2018,regan2017,schiel2019real,ginbo2021,gersonius2011failure,munoz2011,chesney2017,mac2020,truong2016,truong2017managing}. However, to the best of our knowledge, there has not been a study yet that uses real option theory to determine optimal adaptation pathways.

In this paper, we bridge the gap between qualitative and quantitative adaptation pathways by proposing a framework based on real option analysis, which enables the identification of optimal adaptation pathways in an objective manner. \textcolor{black}{Our modeling framework allows to determine whether to invest in low cost adaptation measures first and preserve investment flexibility or to commit to more effective but also more costly adaptation measures first. As illustrated in our numerical application, the answer to this research question is not always straightforwards, and it is important to conduct a quantitative examination of the adaptation problem.}  \textcolor{black}{Our model is built upon and substantially extend previous flood studies by adding real options analysis, especially the analysis with respect to investment sequences. }Firstly, following \cite{menendez2010, lobeto2018, salas2018} we model extreme sea level events, using the block maxima approach from extreme value theory. The block maxima approach is a popular modelling tool to overcome data scarcity often present when estimating catastrophic risk. Secondly, following \cite{aerts2014evaluating}, we adopt the estimation of a flood loss curve in order to translate extreme water levels into monetary losses. Then, similar to \cite{gersonius2011failure, gersonius2013climate} we factor in both the impact and uncertainty of climate change by allowing the mean sea level to follow an arithmetic Brownian motion. Finally, we develop a real option framework to evaluate optimal investment sequences, when several adaptation projects are considered. Therefore, the proposed framework allows us to provide optimal pathways for climate adaptation projects, taking into account catastrophic risk and climate change uncertainty. 

 In applying the model to a case study of NYC, we find that some soft adaptation measures may be preferable to invest immediately when considered separately from other possible measures. However, when all measures are considered together to develop adaptation pathways, it may be optimal to invest in hard measures first. Therefore, one of the key findings of our study is the importance of considering adaptation pathways to determine the best adaptation strategies for coastal cities. Our study extends the literature on real option analysis applied to climate change problems and in particular to flood risk mitigation and adaptation projects \cite[see, for example,][]{gersonius2011failure,gersonius2013climate,woodward2011,woodward2014,kim2017using,brown2018,kim2018,kim2019, dittrich2019making,wreford2020}. We also make important contributions to the literature by extending the case study presented in \cite{aerts2014evaluating}, and linking it to the stream of literature on adaptation pathways \cite[see, among others,][]{downing2012,haasnoot2012,haasnoot2013,wise2014, fazey2016}. Moreover, we extend the discussion on the application of extreme value theory to flood risk modeling and hydrology problems \cite{menendez2010,salas2014,lobeto2018, hieronymus2020, muis2018}. 

 \textcolor{black}{To the best of our knowledge, our study is one of the first to propose a quantitative analysis of climate adaptation pathways based on real option analysis and extreme value theory. Typically, studies on adaptation pathways \citep{downing2012,haasnoot2012,haasnoot2013,wise2014, fazey2016} are more likely to follow a qualitative approach as it comes to modeling flexibility, possibly making the selected pathway and sequence of projects more subjective. In comparison to these applications, in our case study we analyse adaptation pathways based on an underlying quantitative framework, to
overcome the gap between adaptation pathways, optimal timing and the quantitative
analysis for adaptation projects.}

\textcolor{black}{Thus, unlike most of the existing literature, the proposed methodology allows us to study optimal adaptation pathways comprising both soft and hard adaptation
measures. Putting more emphasis on uncertainty and flexibility than existing studies, the application of our adaptation framework also allows us to factor in uncertainty about climate change impacts or growth in the value of the exposure. These features clearly distinguish the work conducted in this study from previous literature on adaptation to climate change impacts.}

The remainder of this paper is organized as follows: Section~\ref{sec:2} presents the modelling framework, Section~\ref{nyc} describes the NYC case study set up, Section~\ref{numerical} shows the empirical results, and Section~\ref{conc} concludes.

\section{Methodology}
\label{sec:2}
This section illustrates the proposed methodology that allows to combine the modeling of losses from flooding with real options analysis for the optimal sequencing of adaptation projects. The proposed framework has three main components: 1) modeling extreme sea water levels, 2) modeling losses from flood risk, and 3) conducting real option analysis for investment sequencing The first component provides for each future time period a distribution of extreme water mark level -- the highest level to which the sea water level rises in a particular year. The second component relates the high water mark level to flood damage via a loss curve. This model calculates the losses without flood mitigation as well as with flood mitigation measures in place at a given time. Finally, the third component uses the reduction in flood losses from a mitigation measure to calculate the value of implementing that measure. This approach allows us to determine the optimal time to implement a single project, or the sequencing of multiple projects when several flood mitigation investment options are available.

\subsection{Sea water level modeling}
Typically, extreme water levels can be modelled using the block maxima approach from extreme value theory. Let $X_1,\dots,X_n$ be independent and identically distributed random variables defined on a common probability space $\left(\Omega,\mathcal{F},\mathbb{P}\right)$, with cumulative distribution function $F$. These can be hourly water levels observed over a time period such as a month or a year. Let $S_t$ be the highest level recorded over the period $t$, i.e. $S_t = \max\left(X_1,\dots,X_n\right)$. According to the Fisher-Tippett-Gnedenko Theorem, the maximum water level distribution converges to the generalized extreme value (GEV) distribution, under some regularity conditions on the cumulative distribution function $F$  \cite[][]{fisher1928limiting}:
\begin{align}\label{gev}
	H(S_{t};m,s,\xi)=
	\exp\left\{-\left[1+\xi \big(\frac{S_{t}-m}{s}\big)\right]^{-1/\xi}\right\}, \quad 1+\xi \big(\frac{S_{t}-m}{s}\big)>0,
\end{align}
where $m, s, \xi$ are called the location, scale and shape parameters. When there are $N$ periods for which observations of the sea level are available, and $S_1,\dots,S_N$ are the block maxima in these periods, the log-likelihood function is given as:
\begin{equation}
    l\left(m, s, \xi,S_1,\dots,S_N\right) = log\left(\prod_{i}^N h(S_{i};m,s,\xi)\mathcal{I}_{1+\frac{\xi(S_i-m)}{\sigma}>0}\right)\notag,
\end{equation}
where $h(S_{i};m,s,\xi)$ is the density function of the GEV defined in Equation (\ref{gev}), and $\mathcal{I}_{x>0}$ is the indicator function being equal to 1 if $x>0$, and zero otherwise.
We define the high water mark level $M_{t}$ as the sum of mean sea-level $W(t)$, the tide $Tide(t)$ and the block maximum surge $S_{t}$. Since $S_{t}$ follows a GEV distribution $H(S_{t};m,s,\xi)$, conditional on $W(t)$ and $Tide(t)$ at time $t$, the high water mark $M_{t}$ also follows a GEV $H(M_{t};\alpha(t),s,\xi)$ with the same scale and shape parameters as $S_{t}$, while the location parameter is given by $\alpha(t) = W(t) + Tide(t)+ m$.

\subsection{Flood loss modeling}

Flood loss curves represent damage to buildings and infrastructures due to inundations, and are regarded as essential components of flood risk assessments \cite[see, for example][]{prettenthaler2010estimation, merz2004estimation}. Nonetheless, the relationship between high sea levels and the consequent destruction that such sea levels can cause is complex and requires careful considerations \cite{hallegatte2011assessing}. Firstly, floodwater loss can be divided into two components: direct and indirect. Direct losses are caused by contact with floodwater, for example damage to buildings, building contents, and infrastructures. Indirect losses are usually considered as a consequence of contact with floodwater, for example business interruption, or the impact of severe flood events on the regional and national economy \cite[see,][]{prettenthaler2010estimation, buchele2006flood, parker2007new}. Secondly, the estimation of a flood loss curve $L(M)$ requires a measurement of the monetary damages due to flood events, which often relies on complex hydrological models \cite[see, ][]{prettenthaler2010estimation, hallegatte2011assessing, aerts2014evaluating, wang2015adaptation, han2020agent}. In the case study, we therefore adopt the loss curve estimated by \cite{aerts2014evaluating} and assume that it is available to the decision-maker.

Given a loss curve $L(M)$, we can estimate future flood losses once we have made some assumptions about the growth of the loss exposure through time. When the loss exposure grows at a rate $\gamma$, a water level $M_{t}$ will cause a loss $e^{\gamma t}L(M_t)$. Adaptation projects help reduce flood losses and the value they provide over a future period $t$ is likely to be higher than the expected reduced flood losses since reduced flood risk leads to more stable wealth for affected residents. In cases that all properties and infrastructure in the area are fully insured, the benefit provided by adaptation projects in a period $t$ is the reduction in insurance premium for that period. When the mean sea-level is $\alpha$, the insurance premium for a full loss coverage in period time $t$ can be obtained as
\begin{align}\label{premium}
	\pi(u,t,\alpha;L) = (1+\delta)e^{\gamma t}\int_{u}^{\infty}L(M)dH(M),
\end{align}
where $\delta$ is the premium loading and $u$ is the flood threshold such that when the water level is lower than $u$, no damage is caused\footnote{\textcolor{black}{See the appendix for more details.}}. \\

The impact of adaptation projects on flood losses can also be analyzed in this framework. In the following we consider two competing adaptation measures: flood-proofing of buildings and the construction of a barrier and dike system. Flood-proofing is a soft climate adaptation measure, as it is usually less invasive and often limited only to building-specific improvements. Common examples of flood-proofing are the installation of water shields on windows, reinforcements to external walls, and the usage of specific sealants to reduce the impact of water through walls.

\begin{figure}[h!]
	\begin{center}
		\includegraphics[scale=0.35,angle=0]{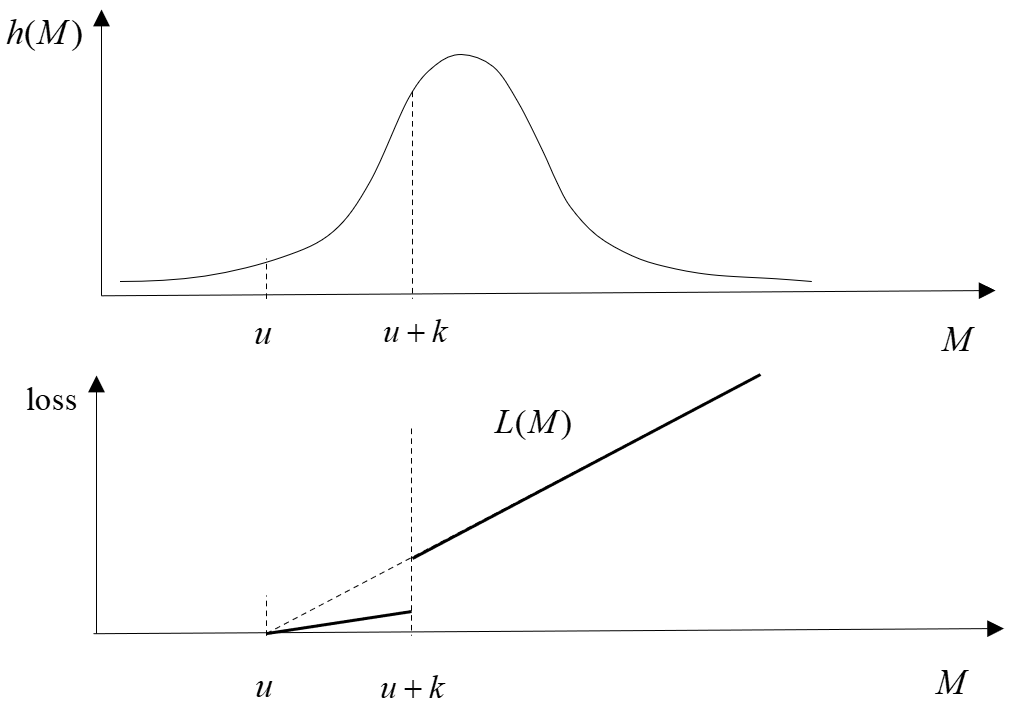}
		\caption{This figure shows an example for the density of the GEV distribution (top), and the impact of flood-proofing measures on the loss curve (bottom). Flood proofing measures reduce the impact of sea levels between $u$ and $u+k$, by rotating the loss curve of the proportion $\kappa$.}
		\label{proofing}
	\end{center}
\end{figure}

Flood-proofing can help to reduce damages by a proportion $\kappa$ as long as the water level is lower than a certain height, e.g., $u+k$. Figure \ref{proofing} shows the impact of flood-proofing measures on the loss curve $L(M)$. Adopting flood-proofing measures, only affects the loss curve for sea level between $u$ and $u+k$ by rotating downwards the curve of the proportion $\kappa$. By alleviating the impact of smaller floods, flood-proofing measures reduce the flood losses by the following amount:
\begin{align}
	R(k,t,\alpha;L) = \kappa [\pi(u,t,\alpha;L) - \pi(u+k,t,\alpha;L)].
\end{align}
The construction of a dike and a barrier aims to protect an entire area from the sea level rise up to a given height. It is considered to be a hard climate adaptation measure, and it requires major investment by the decision maker. Building a dike or seawall that raises the flood threshold from $u$ to $u+k$ will help avoid damages from floods with levels up to $u+k$. The impact is similar to that of flood-proofing where $\kappa = 1$. The dike helps to reduce the flood loss at time $t$ from $\pi(u,t,\alpha;L)$ to $\pi(u+k,t,\alpha;L)$, and so the benefit flow of this adaptation measure is
\begin{align}
	R(k,t,\alpha;L) = \pi(u,t,\alpha;L) - \pi(u+k,t,\alpha;L).
\end{align}

The flood loss curve in \cite{aerts2014evaluating} can be used to translate the level of the flood into monetary losses. Nevertheless, \cite{aerts2014evaluating} only provides discrete estimates for the flood loss curve. Therefore, to approximate the true relationship between monetary losses and floodwater level, we employ a polynomial curve.
In empirical applications, polynomial curves can be used to describe the relationship between extreme water levels and monetary losses from flooding. Often a quadratic loss function $L(M)$ is used and we also follow this approach in our study \cite[see, for example][]{merz2004estimation,merz2010,seifert2010,kreibich2010, gerl2016},
\begin{align}\label{quad1}
	L(M)=aM^2+bM+c=X\beta,
\end{align}
where $X = [\begin{array}{ccc} M^{2} &M&1\end{array}]$ and $\beta =[\begin{array}{ccc} a &b&c\end{array}]^{'}$\footnote{\textcolor{black}{See the appendix for more details.}}. Under this specification the implicit flood threshold, i.e. the sea level corresponding to a loss of zero can be found as $u^*$  such that $a(u^*)^2 + bu^* + c=0$.  We consider $u^*$ as the flood threshold in absence of any flood mitigation measure. With the quadratic loss curve, the expected damage from flooding at a general flood threshold $x$, $D(x,\alpha;\beta)=\int_{x}^{\infty}L(M)dH(M)$, can be determined in closed forms. In the case of the GEV with the cumulative distribution function as in Equation (\ref{gev}), the expected damage function $D(x, \alpha;\beta)$ has the following form \citep{el2013parameters,embrechts2013modelling}:
\begin{align}\label{Dxa}
	D(x, \alpha;\beta)= \left\{\begin{array}{ll}
		D_{1}(x, \alpha;\beta) & \text{if } \xi>0 \text{ and }x>\alpha - \frac{s}{\xi} \\
		D_{2}(x, \alpha;\beta) & \text{if } \xi>0 \text{ and } x\le \alpha - \frac{s}{\xi} \\
		0 & \text{if } \xi<0 \text{ and }x>\alpha - \frac{s}{\xi} \\
		D_{1}(x, \alpha;\beta) & \text{if } \xi<0 \text{ and }x\le\alpha - \frac{s}{\xi}
	\end{array}\right.
\end{align}
where $D_{1}(x, \alpha;\beta)$ and $D_{2}(x, \alpha;\beta)$ are given by:
\begin{align} \label{dxtext}
	D_{1}(x, \alpha;\beta)&=a(\frac{s}{\xi})^{^2}\Gamma\big(1-2\xi,\big[1+\xi \big(\frac{x-\alpha}{s}\big)\big]^{-1/\xi}\big)  \nonumber \\
	&+2\frac{s}{\xi}(a\alpha - a\frac{s}{\xi} + b/2)\Gamma\big(1-\xi,\big[1+\xi \big(\frac{x-\alpha}{s}\big)\big]^{-1/\xi}\big)  \nonumber \\
	&+[a(\alpha - \frac{s}{\xi})^{2} +b(\alpha - \frac{s}{\xi}) + c]\big[1-H(x)], \\
	D_{2}(x, \alpha;\beta)&=
	a(\alpha-\frac{s}{\xi})^{2} + \frac{sa}{\xi}\bigg(\frac{s}{\xi}\Gamma(1-2\xi)-2(\frac{s}{\xi}-\alpha)\Gamma(1-\xi)\bigg)\nonumber\\
	&+b\alpha + \frac{sb}{\xi}[\Gamma(1-\xi)-1] + c.
\end{align}

\subsection{Real option model} \label{inv}
Recall that $M_{t}$ follows a GEV distribution  $H(M_{t};\alpha(t),s(t),\xi)$, where $\alpha(t)= W(t) + Tide(t)+ m$. Tide presents seasonality effects that can be challenging to deal with in investment analysis. We simplify the investment model by adopting the deseasonalization approach often used in the real option literature, see e.g. \cite{abadie2008european,boomsma2012renewable,ernstsen2018valuation} and in flood risk adaptation, see e.g. \cite{aerts2013low, aerts2014evaluating, wang2015adaptation, hallegatte2011assessing}.

Climate change uncertainty is typically modeled via a Brownian motion \cite[see, for example][]{gersonius2011failure,gersonius2013climate, kim2019}. Here we assume that the mean sea level $W(t)$ follows the following arithmetic Brownian motion process\footnote{More complex formulations for the mean sea level process can be adopted, such as, e.g., the introduction of a feedback loop, which is left for future research.}: 
\begin{align}\label{slr}
	dW(t) = \mu dt + \sigma dB(t),
\end{align}
where $B(t)$ is a standard Brownian motion, $\mu$ is the expected change in the mean sea-level and $\sigma$ represents the uncertainty in the estimates of sea-level changes. Representing the mean sea level evolution as an additive process of random shocks closely resembles the long term trend behavior of extreme sea level events, as often found in applications of extreme value theory on extreme sea level events \citep{menendez2010,lobeto2018,salas2018}. Then, the location parameter $\alpha(t)$ also follows an arithmetic Brownian motion
\begin{align}
	d\alpha(t) = \mu dt + \sigma dB(t),
\end{align}
and the location parameter for the high water mark at time $T$ is $\alpha(T) \sim N(\alpha_{0} + \mu T, \sigma^{2}T)$. The simple relation between the location parameter $\alpha_{t}$ and the mean water level provides a way to estimate the time varying location parameter $\alpha_{t}$ by observing the mean water level. We estimate the mean water level by the average hourly water level for each year, and use the annual mean water level to estimate the stochastic process in Equation (\ref{slr}).

In order to evaluate a climate adaptation project, we need a risk neutral probability measure, which can be obtained using the  Capital Asset Pricing Model (CAPM). According to the CAPM, the expected return $\mu_{X}$ on an asset $X$ that has a return standard deviation $\sigma_{X}$ is given by:
\begin{align}
	\mu_{X} = r + \phi \rho_{xm} \sigma_{X},
\end{align}
where $r$ is the risk free rate, $\rho_{xm}$ is the correlation between the asset return and market return, $\phi = \frac{r_{m}-r}{\sigma_{m}}$ is the market price of risk and $r_{m}$ and $\sigma_{m}$ are the expected value and standard deviation of the market return \cite[see, e.g.,][]{sharpe1964capital, lintner1964optimal,mossin1966equilibrium}. Under the risk neutral measure, the rate of return of asset $X$ is then $r$ or $\mu_{X} - \phi \rho_{xm} \sigma_{X}$. \\
Using similar lines of argument, we can then define the risk neutral process for the location parameter $\alpha(t)$ as:
\begin{align}
	d\alpha(t) &= (\mu^{} - \theta)dt + \sigma dB^{*}(t),
\end{align}
where $\theta = \phi \sigma\rho_{wm}$ and $\rho_{wm}$ is the correlation between the changes in the mean sea level and the market return, and $dB^*(t)= dB(t) +  \phi \rho_{xm} \sigma dt$ is a Brownian motion under the risk neutral measure $\mathbb{P}^{*}$. Using the risk neutral measure, the project value can then be calculated as
\begin{align}
	V(\alpha) = \int_{t}^{\infty}e^{-(r-\gamma)(u-t)} E^* \left[R(k,\alpha(u)) \middle\vert \alpha(t) = \alpha \right]du,
\end{align}
and the value of the option to invest, $\Phi(\alpha)$, can be shown to follow the differential equation, see, e.g., \cite{Dixit1994}:
\begin{align} \label{sdemf}
	\frac{1}{2}\sigma^{2}\Phi_{\alpha \alpha} +(\mu-\theta) \Phi_{\alpha} - r\Phi = 0.
\end{align}
where $\Phi_\alpha$ and $\Phi_{\alpha\alpha}$ are the first and second derivative of $\Phi$, respectively. Furthermore, at the option exercise threshold $\alpha^{*}$, the value matching and smooth pasting conditions need to be satisfied, providing two boundary conditions:
\begin{align}
	\Phi_{}(\alpha^{*})&=V(\alpha^{*}) - I, \\
	\Phi_{\alpha}(\alpha^{*})&=V_{\alpha}(\alpha^{*}),
\end{align}
where $I$ is the investment cost of the project.

When several adaptation projects can be invested, investment sequencing can be important. We define the optimal investment sequence as the one that gives the highest investment value. For an example of two mutually non-exclusive investment projects, Project 1 and Project 2, two investment sequences are possible: i) Project 1 is invested first, and then Project 2 is invested some time after and ii) Project 2 is invested first and then Project 1 is invested. It is necessary to compute the investment value for each sequence and identify the optimal sequence that gives the highest value. The same procedure applies when more than two projects are considered.

To evaluate the investment value of one investment sequence, for example, the first sequence in the above example, suppose the investment costs of the two projects are $I_{1}$ and $I_{2}$. In addition, suppose $R_{1}(k_{1},t,\alpha;L_{1})$ is the benefit flow that is obtained when Project 1 is invested. The benefit flow $R_{1}(k_{1},t,\alpha;L_{1})$ is the same as in the case of a single investment only. In contrast, the benefit flow $R_{12}(k_{2},t,\alpha;L_{2})$ obtained when we further invest in Project 2, after Project 1 has been invested, is the marginal benefit obtained due to the additional protection offered by Project 2.  For example if we invest in flood-proofing measures first, and in the construction of the dike at a later stage, the benefit flows can be written as:
\begin{align}
    R_1(k_1,t,\alpha;L) &= e^{\gamma t} \kappa(1+\delta) \left(D(u,\alpha;\beta) - D(u+k_1,\alpha;\beta)\right)\notag\\
    & = \kappa\left(\pi(u,t,\alpha;L) - \pi(u+k_1,t,\alpha;L)\right)\notag
\end{align}
and
\begin{align}
    R_{12}(k_2,t,\alpha;L) &= e^{\gamma t}(1+\delta)\left(D(u+k_1,\alpha;\beta) -D(u+k_1+k_2,\alpha;\beta)\right)\notag\\
    & = \pi(u+k_1,t,\alpha;L) - \pi(u+k_1+ k_2,t,\alpha;L).\notag
\end{align}
Similarly, if the dike and barrier project is adopted first, and the flood-proofing measures are applied afterwards, the benefit flows can be written as:
\begin{align*}
    R_2(k_2,t,\alpha;L) &= e^{\gamma t}(1+\delta) \left(D(u,\alpha;\beta) - D(u+k_2,\alpha;\beta) \right) \notag\\
    & = \pi(u,t,\alpha;L) - \pi(u+k_2,t,\alpha;L)\notag
\end{align*}
and
\begin{align}
    R_{21}(k_1,t,\alpha;L) &=  e^{\gamma t} \kappa(1+\delta) \left(D(u+k_2,\alpha;\beta) - D(u+k_2+k_1,\alpha;\beta)\right)\notag\\
    & = \kappa\left(\pi(u+k_2,t,\alpha;L) - \pi(u+k_2+k_1,t,\alpha;L)\right).\notag
\end{align}

Then, to determine the optimal time $\tau_{1}$ to invest in Project 1, and the optimal time $\tau_{2}$ to invest in Project 2 thereafter, the following investment problem needs to be considered:
\begin{align}
	\Phi_{1}\left(\alpha(t)\right)+\Phi_{12}\left(\alpha(t)\right)=& \max_{\tau_{1},\tau_{2} } E^{\ast}_{t} \left[ \int_{\tau_{1}}^{\infty} e^{-rs}R_{1}(k_{1},t,\alpha(s);L_{1})ds-e^{-r\tau_{1}}I_{1} \right. +  \\
	&\qquad \quad \left. \int_{\tau_{2}}^{\infty}e^{-rs}R_{12}(k_{2},t,\alpha(s);L_{2})ds-e^{-r\tau_{2}}I_{2}\middle\vert \alpha(t) \right]. \nonumber
\end{align}
Hereby, $\Phi_{1}(\alpha(t))$ is the value of the option to invest in Project 1 and $\Phi_{12}(\alpha(t))$ is the value of the option to invest in Project 2, after Project 1 has already been invested. Similarly, if Project 2 is invested first, then $\Phi_{2}(\alpha(t))+\Phi_{21}(\alpha(t))$ is obtained. In considering which project to be invested first, the decision maker needs to select a sequence that maximizes the option to invest, $\Phi_{s}(\alpha(t))$, i.e.
\begin{align}
\label{dynamic}
	\Phi_{s}(\alpha_{t})=\max\{\Phi_{1}(\alpha(t))+\Phi_{12}(\alpha(t)), \Phi_{2}(\alpha(t))+\Phi_{21}(\alpha(t)) \} .
\end{align}

To find the project values, optimal investment boundaries, and option values for the problem, we use the binomial lattice to solve for the project value, the optimal investment boundary and the option value, since it is simple to find the investment boundary with this method.

\section{Case Study}
\label{nyc}
In this section we present a case study to demonstrate the application of our proposed methodology to assess optimal adaptation pathways. Consider a decision maker, willing to invest into climate adaptation, aiming to protect NYC from the risks of sea level rise and extreme flooding. The decision maker is tasked to evaluate the opportunity to invest into one or more admissible climate adaptation projects. The set of admissible projects is composed by: i) a barrier and dike project and ii) a flood proofing project. Note that the framework presented in the previous section can be easily extended to analyse an investment in a portfolio of projects composed of more than two alternatives.

The barrier project is referred to as project S2c  in \cite{aerts2014evaluating}, which aims to reduce the length of the coastline of the NYC area as much as possible. It involves the construction of two storm surge barriers. One barrier will connect Sandy Hook in New Jersey and the tip of the Rockaways in Queens, New York and the other barrier will close the East River. The investment cost is estimated to range from \$11 billion to \$15 billion and the maintenance cost is \$118 million per year. We use a mid-range value of the investment cost, which is \$13 billion. The project raises the flood threshold by 1000mm for houses and infrastructure.

The flood-proofing project is a measure that helps to keep the options open. Building codes can be further enhanced in the future or storm surge barriers can be developed. It involves wet flood-proofing of existing buildings in the one-in-100 year flood area. We consider the most cost effective measure found by \cite{aerts2014evaluating}, which involves wet-proofing 2 feet (610mm) of buildings with 30\% effectiveness. The cost of this measure is \$246 million.

\subsection{Sea level data}
Data for the empirical studies in this project are collected  from the archives of the University of Hawaii Sea Level Centre (\url{http://uhslc.soest.hawaii.edu}). The sea level time series is composed of hourly observations of sea level taken at the gauge station located at $40.7^{\circ}$ North $74.015^{\circ}$ West, and contains hourly observation of the sea level from 6 February 1920 until 31 December 2019. This time series has been used in many previous studies, see \textit{e.g.} \cite{menendez2010}. We use the most recent 4 years to estimate the tide using the \textit{UTide} package by \cite{codiga2011unified}.
\begin{figure}[h!]
	\begin{center}
		\includegraphics[scale=0.7,angle=0]{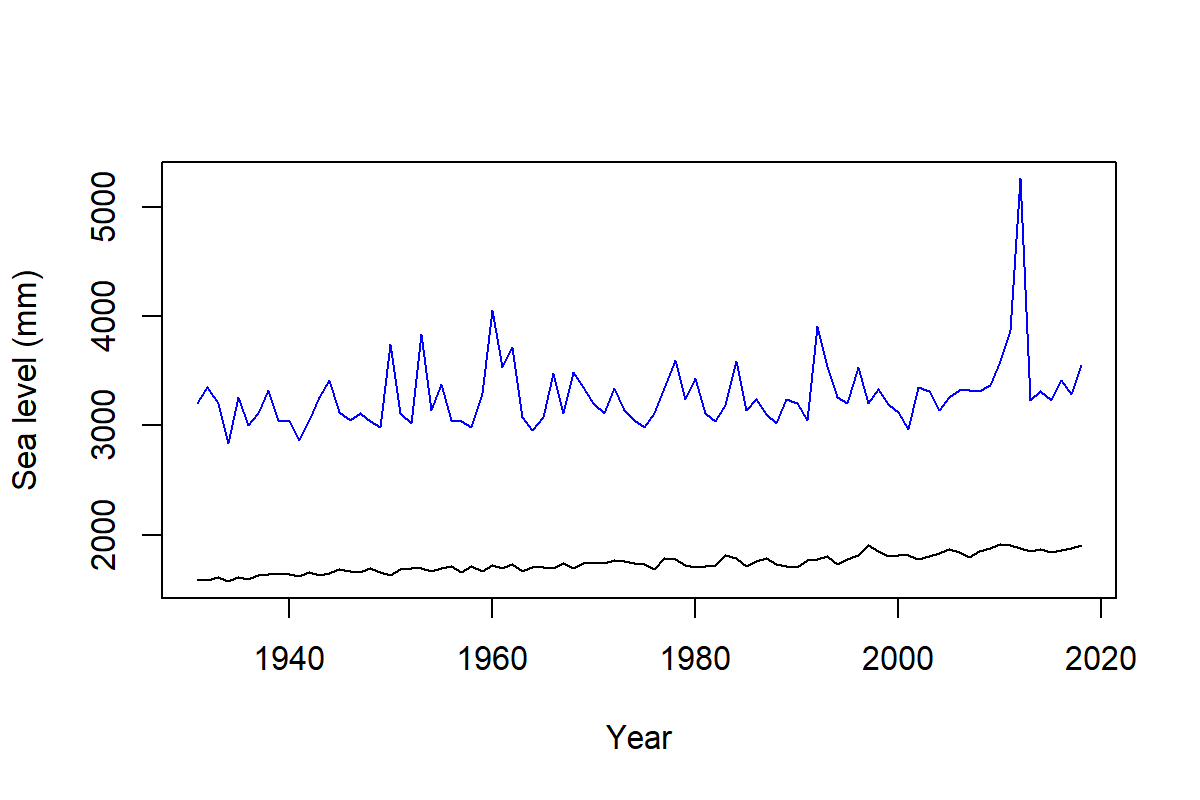}
		\caption{Observed annual mean and annual maximum sea levels at the Battery in lower Manhattan, NYC.}
		\label{nysl}
	\end{center}
\end{figure}
Figure \ref{nysl} shows the annual mean (black line) and annual maximum (blue line) sea level at the NYC gauge station between  February 1920 and December 2019. We observe that both the annual mean and annual maximum exhibit an increasing trend, while the annual maximum exhibits a much higher volatility than the annual mean.

\subsection{Adaptation projects}

We examine the optimal investment strategies for two projects for NYC based on \cite{aerts2014evaluating}: a) a barrier project that uses barriers and dikes to protect the region and b) a wet flood-proofing project that adjusts the interior and material of buildings in the one-in-100 year flood area (i.e. the area in which the probability of having a flood in any single year is at least 0.01) to limit flood damage.
\begin{table}[]
	\caption{Information on estimated and assumed parameter values.}
	\label{estimate}
	\centering
	\begin{tabular}{p{2.5 in}p{1in}} \hline
		Parameters & Values \\ \hline
		Location parameter ($\alpha_0$) & 1642 \\
		Scale parameter ($s$) & 131 \\
		Shape parameter ($\xi$) & 0.27 \\	
		Current flood threshold ($u$) & 2506mm \\
		Quadratic coefficient of damage curve ($a$) &  0 \\
		Linear coefficient of damage curve ($b$) &0.0393 \\
		Mean sea level rise per year ($\mu$) & 6mm \\
		Standard deviation of sea level rise per year ($\sigma$) & 25mm \\
		Risk free rate ($r$) & 4\% \\
		Growth rate of loss exposure ($\gamma$) & 1\% \\
		Market risk premium ($\theta$) & 0.15 \\
		\hline
		\textit{Flood-proofing (Project 1)}&\\
		Investment cost ($I_{1}$) & \$246 million \\
		Flood mitigation ($k_{1}$) & 610mm	\\	
		Effectiveness ($\kappa$) & 30\%	\\	
		\hline
		\textit{Dike construction (Project 2)}&\\
		Investment cost ($I_{2}$) & \$15.95 billion\\
		Flood mitigation  ($k_{2}$) & 1000mm \\
		\hline
	\end{tabular}
\end{table}

We start by estimating flood risk in the area based on the surge height at the Battery in lower Manhattan, NYC. We obtain the surge height by excluding the tide and the mean sea level from the hourly water level. We then obtain the annual maximum of storm surge height for each year to estimate the GEV distribution. The average tide and the mean sea level are then added back to the estimated location parameter to obtain the distribution of the high water mark. The estimated parameters of the high water mark distribution are: $\alpha=1642, s = 131, \xi = 0.27$. We assume a 1.31\% growth in loss exposure for the area in this study, based on GDP and population growth in the region (see, \cite{weinkle2018normalized}, for details). In addition, we assume a real risk-free rate of 4\% based on the study by \cite{newell2003discounting}, who estimate the real risk-free rate using 10-year treasury bond yields observed since 1800.\footnote{This is also the discount rate used by \cite{aerts2014evaluating}.} We use an estimate of the market risk premium $\theta$ of 0.15, based on the correlation between the annual change in the mean sea level and the rate of return on the stock market index (S\&P 500) (corr= 0.0283), the market price of risk $\phi = 0.1382$, and the historical standard deviation of 38.48. A safety loading of $\delta = 3\%$ is assumed, following the estimation results reported by \cite{shao2015catastrophe} based on CAT bond prices. We use results from climate change studies to calibrate the mean sea level rise process. We follow \cite{aerts2014evaluating} to assume a mean sea level rise of 6mm per year for the area and use $\sigma=$25mm to summarise the uncertainty in sea level change estimates, based on data provided by \cite{johansson2014global}. The ratio of the standard deviation and the mean of the sea level rise is quite high, which indicates significant uncertainty related to the estimate of future sea level changes.

\begin{figure}[]
	\begin{center}
		\includegraphics[scale=0.65,angle=0]{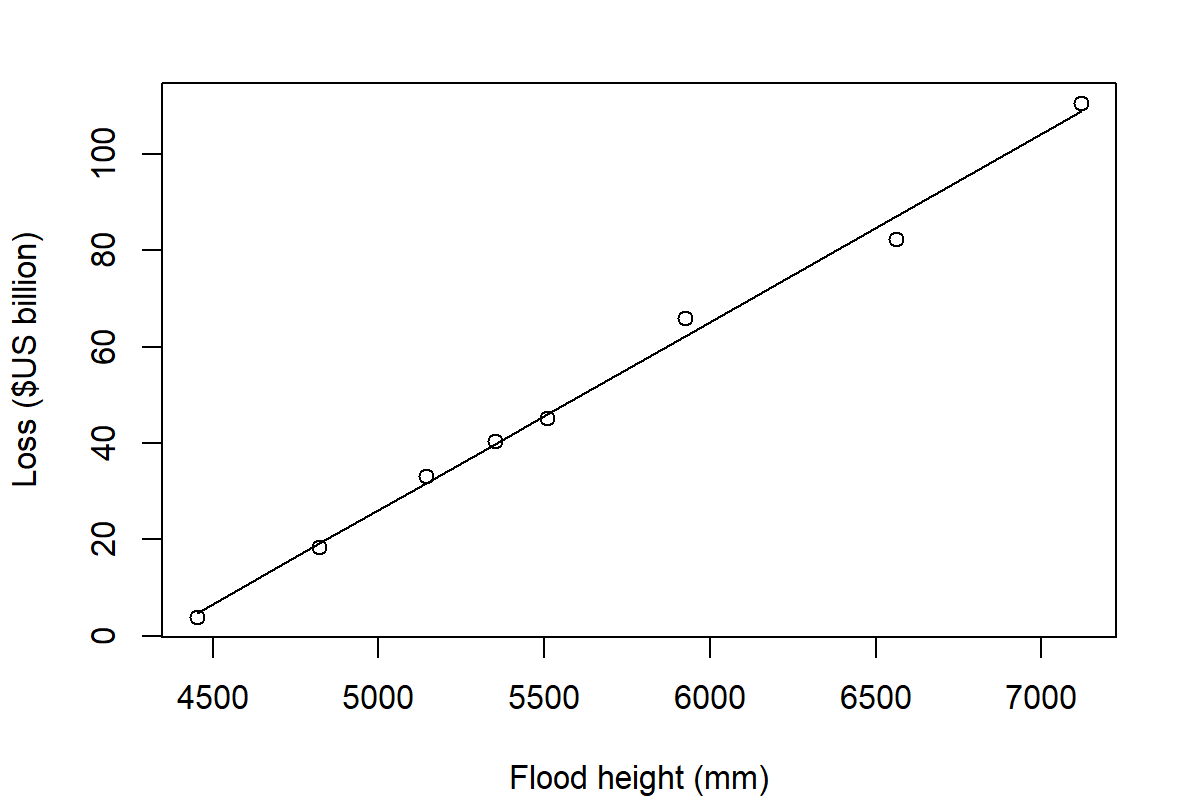}
		\caption{Damage curve for the studied area in New York City.}
		\label{damagecurve}
	\end{center}
\end{figure}

We adopt the damage curve estimated by \cite{aerts2014evaluating} (Figure \ref{damagecurve}). Without adaptation, flood loss occurs whenever the sea level exceeds 2.506 metres. This flood threshold occurs at the extreme right tail of the distribution where $1-H(u)=0.023$. Using the estimated damage curve and the high water mark distribution, the expected annual damage without adaptation is \$498M, which is slightly lower than the expected annual damage of \$508M estimated by \cite{aerts2014evaluating}. The discrepancy between the two estimates is likely due to the difference in the approach to high water mark estimation: \cite{aerts2014evaluating} use a model-based approach that estimates the high water mark distribution based on simulated storm surge events, while we use historical data.

The flood-proofing project will reduce flood damages to buildings, leaving flood losses to business interruption, infrastructure and vehicles unchanged. \cite{aerts2014evaluating} consider flood-proofing for buildings in  the one-in-100-year flood area, leaving buildings in other areas unchanged. To recognize the impact of this measure in our model, we scale down the total damage curve in Figure \ref{damagecurve} such that the annual benefit of flood-proofing under no growth in the exposure and no climate change is the same as that estimated by \cite{aerts2014evaluating}.

\begin{figure}[h!]
	\begin{center}
		\includegraphics[scale=0.65,angle=0]{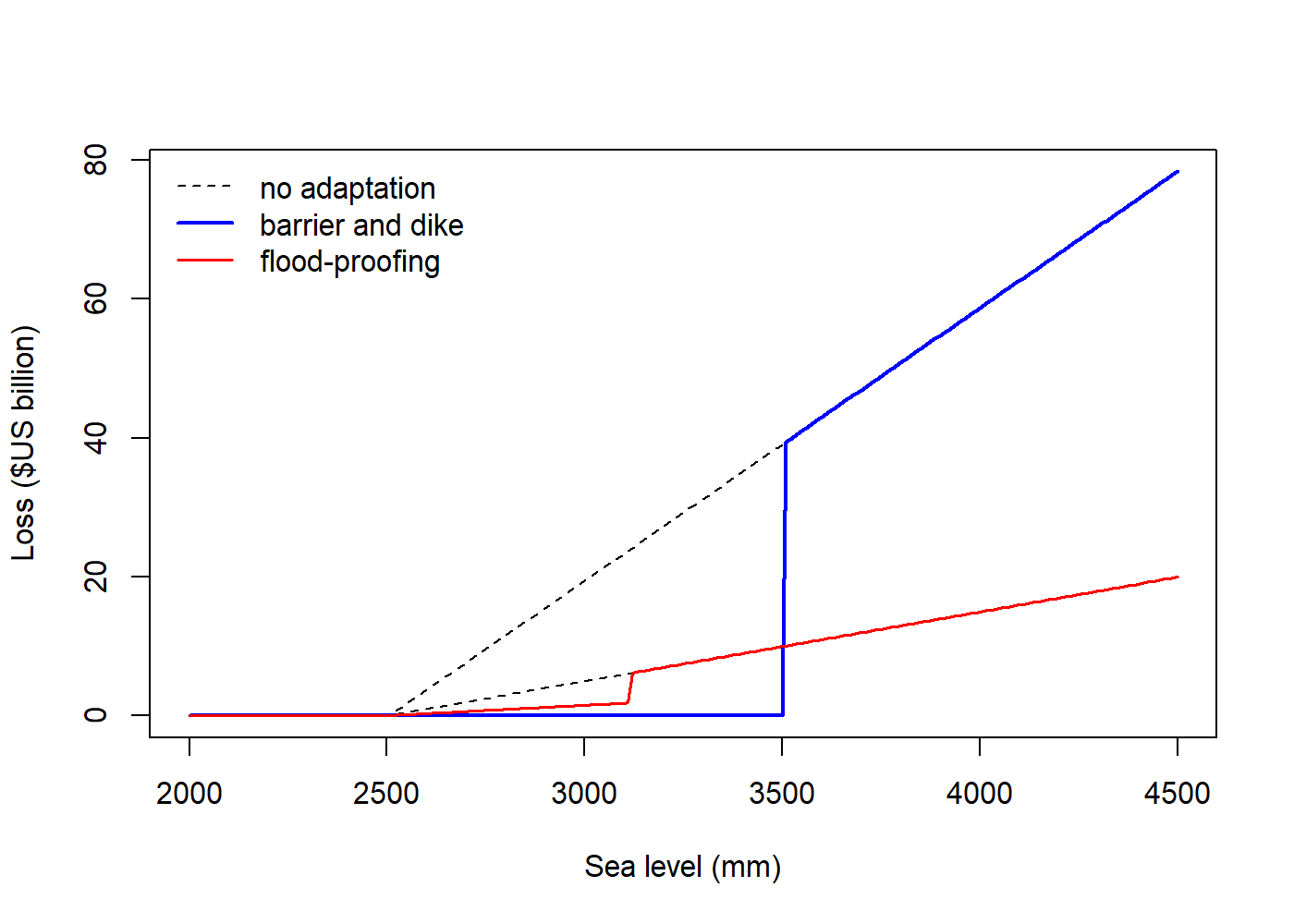}
		\caption{Damage curves estimated for the case of no adaptation, adaptation with barriers and dikes and adaptation with flood-proofing buildings.
		}
		\label{damagecurves_nyc}
	\end{center}
\end{figure}

Figure \ref{damagecurves_nyc} shows the damage curves for the two projects. The barrier and dike project helps to prevent water from coming into the city and provides a large risk reduction. In contrast, the flood-proofing project focuses on only protecting buildings and leaving business interruptions and damages to infrastructure unchanged. These two projects may provide a good example of supplementary projects in the dynamic dimension. It appears that the barrier project is costly and optimal to invest only when climate change is sufficiently serious. The flood-proofing project has a low investment cost and may be worthwhile to invest in the absence of the barrier project. Actual investment timing, however, will depend on the benefit of each project relative to the cost.
\vspace{-0.2in}
\section{Empirical Results}
\label{numerical}
In this section we present the empirical results for the application of our proposed framework to the illustrated case study of optimal adaptation pathways for reducing the risk from sea level rise and extreme flooding for NYC. Firstly, we study the optimal adaptation pathways for the two flood risk mitigation measures in a baseline scenario. Then, we perform a sensitivity analysis to assess the robustness of our numerical results with respect to sea level rise, discount rate, and sea level rise uncertainty.
\subsection{Baseline scenario}
Table \ref{analysis_nyc_portfolio_de} provides the results for two alternative pathways: invest in the flood-proofing measure first and in the barrier and dike measure later (upper panel); ii) invest in the barrier and dike measure first and consider the flood-proofing measure later (bottom panel). We provide the net present value (NPV) and the value of the option for both individual projects, as well as the total value. Looking at the upper panel, it can be seen that flood-proofing returns a positive NPV, with its value being equal to the total value with optionality, i.e. considering the optimal timing of the project through real option analysis. This implies that both a decision rule based on NPV and real option analysis would suggest that the flood-proofing project is invested right away.

For the barrier and dike project, the NPV is positive, but lower than the total value with optimal timing of the project. Thus, while investing right away in the dike and barrier project would provide a positive return, it is actually more beneficial to defer the investment into the future and to wait for the uncertainty about climate change to unfold. This result illustrates the benefits of applying real option analysis in comparison to choosing a simple NPV approach.

In terms of dynamic pathways, investing first in the flood-proofing and then into the barrier and dike project, either right away or in the future, provide a positive total value. The bottom panel presents the dynamic pathways for investing in the dike and barrier project first, and then into the flood-proofing measures. Similar to the previous case, the barrier and dike project exhibits a positive NPV but lower than the total value with optionality, indicating that even when the barrier and dike project is invested first, based on our conducted real option analysis the decision to invest should be deferred into the future. Interestingly, when the flood-proofing project is invested after the barrier and dike, it yields a negative NPV and still a zero benefit with optionality, i.e. when real option analysis is applied. This indicates that under the baseline scenario the barrier and dike project seems to be sufficient to defend NYC from sea level rise and makes additional flood-proofing measures redundant.

Overall, the results for our baseline scenario suggest that it is not optimal to invest right away. The highest project value (USD $10.69$ billion) can be achieved by waiting initially and then investing in the barrier and dike project some time in the future. Our results also suggest that once the barrier and dike project has been invested, there is no additional benefit of investing into the flood-proofing project at a later stage. Interestingly, an alternative pathway, providing a slightly lower benefit (USD $10.38$ billion) is to invest into the flood-proofing project immediately and then into the barrier and dike project at a later stage.

\begin{table}[h]
    \centering
    \caption{Investment analysis for investment sequences using NPV rule and real option methods. We report (in billion USD) the NPV, the total value with optionality and the difference of the two values.}
    \label{analysis_nyc_portfolio_de}
    \begin{tabular}{lrrrrrr}
      \hline
      &\multicolumn{5}{c}{Flood-proofing then Barrier and Dike}\\
       \cmidrule{2-6}
      & Flood-proofing & & Barrier and Dike & &Total\\
      \cmidrule{2-2} \cmidrule{4-4} \cmidrule{6-6}
      NPV & 0.78 &  & 7.63 &  & 8.42\\
      Total value with optionality   & 0.78 &  & 9.60 &  & 10.38\\
      Difference  & 0.00 &  & 1.97 &  & 1.97 \\
      \hline
      &\multicolumn{5}{c}{Barrier and Dike then Flood-proofing}\\
      \cmidrule{2-6}
      & Barrier and Dike  & & Flood-proofing & &Total\\
      \cmidrule{2-2} \cmidrule{4-4} \cmidrule{6-6}
      NPV & 9.15 &  & -0.73 &  & 8.42 \\
      Total value with optionality  & \bf{10.69} &  & 0.00 &  & \bf{10.69} \\
      Difference & 1.54 &  & 0.73 &  & 2.27 \\
      \hline
    \end{tabular}
\end{table}

\begin{figure}[h!]
	\begin{center}
		\includegraphics[scale=0.65,angle=0]{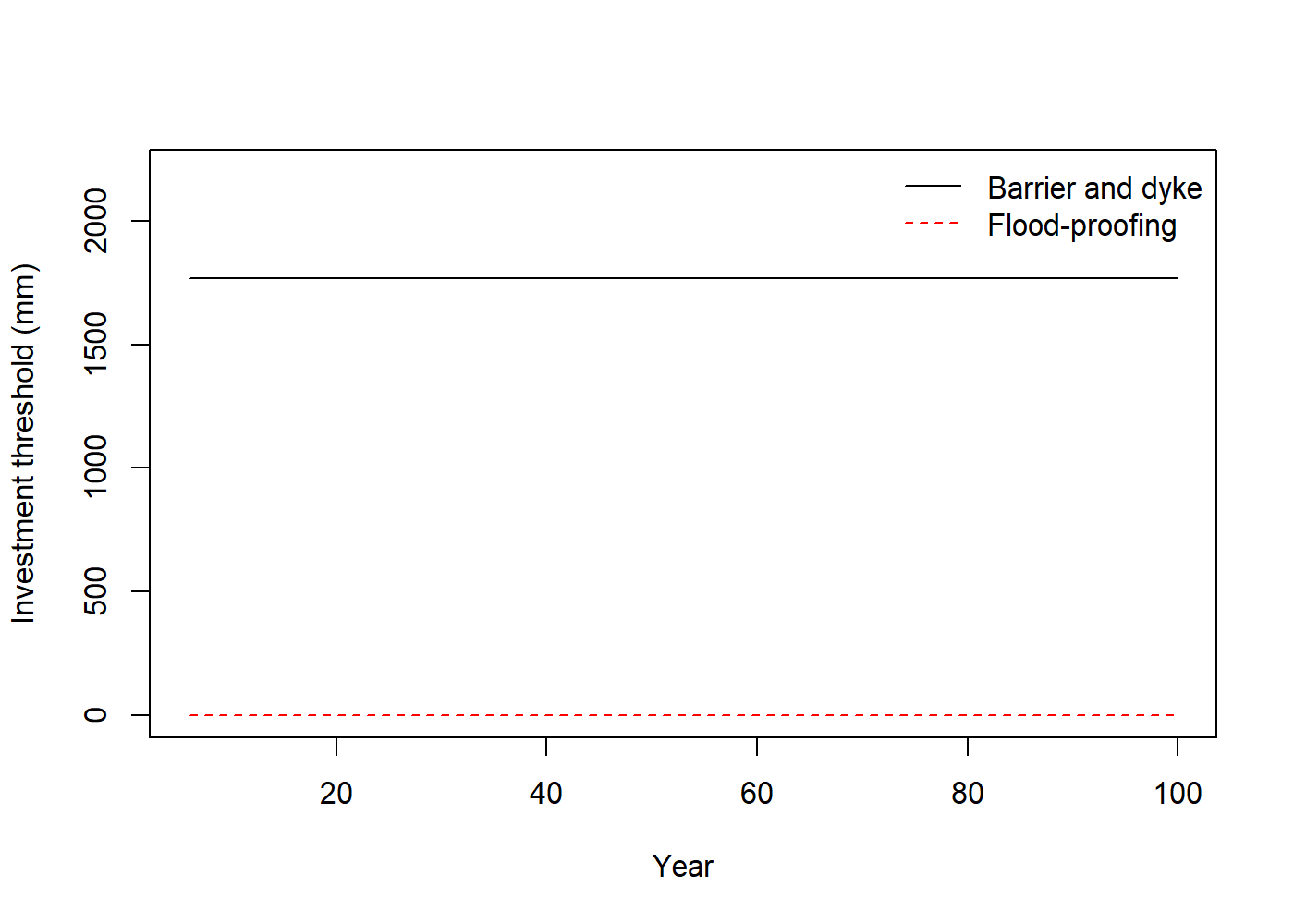}
		\caption{Exercise boundaries (water levels) for the dynamic pathway of investing into the flood-proofing project first and into the barrier and dike project second}
		\label{ex_bounds}
	\end{center}
\end{figure}

Figure~\ref{ex_bounds} shows the exercise boundaries for the conducted real options analysis in terms of water levels for the first dynamic pathway analysed in Table~\ref{analysis_nyc_portfolio_de}, i.e. investment into flood-proofing first and then into the barrier and dike project. The exercise boundaries in Figure~\ref{ex_bounds} confirm our numerical results. When flood-proofing is invested first, the water level required to initiate the adoption of the mitigation measures is close to zero, suggesting that the adoption happens right away. The barrier and dike project is then invested at a significantly higher water level of approximately 1.8 metres. Note that we do not provide a figure for the second dynamic pathway, since once the barrier and dike project has been invested (optimal at an approximate water level of 1.4 metres), there is no need to invest into the flood-proofing project anymore, i.e. the water level required for additional investment would go to infinity.

\subsection{Sensitivity analysis}
In this subsection we present a sensitivity analysis to examine the impact of some of the variables on our results. First, we investigate how robust the results of our numerical analysis are with respect to the discount rate. Table~\ref{sensitivity_portfolio_de_nyc_discount} shows the total investment value for the optimal adaptation pathways under different discount rate regimes. The upper panel shows the results for the case where the flood-proofing measures are adopted first, and then the barrier and dike project is invested, while the bottom panel shows the opposite investment sequence, with the barrier and dike project being invested first, and the flood-proofing measures are adopted subsequently. Note that for each scenario, we indicate the project value of the optimal adaptation pathway in bold.

As expected, climate adaption project values depend significantly on the discount rate, with values of the invested projects decreasing as the discount rate increases. Nevertheless, the strategy to invest in the dike and barrier project first, remains optimal under all discount rates considered, with the exception of $r = 6\%$, where the recommendation is to invest first into the flood proofing project. Note, however, that for $r = 6\%$, the two adaptation pathways are approximately equivalent in terms of project value.

Table \ref{sensitivity_portfolio_de_nyc_slr} shows the sensitivity analysis with respect to different scenarios for mean sea level rise, ranging from $\mu=0$mm up to $\mu=12$mm per year. As the sea level rise increases, so does the project value for both the individual projects and also for both adaptation pathways considered. Interestingly, the pathways selected as optimal in the previous section, (flood-proofing first, barrier and dike then) is optimal for higher levels of sea level rise ($\mu\geq 6$), while for lower levels of sea level rise, investing in the flood-proofing measures first, and then waiting for the climate change uncertainty to unfold and finally invest into the barrier and dike project, returns a higher project value. This important result illustrates that adaptation pathways, i.e. the optimal sequencing of projects, may well be affected by key variables such as the assumed magnitude of sea level rise through time.

Finally, Table~\ref{sensitivity_portfolio_de_nyc_sigmas} shows the results of the conducted sensitivity analysis for different scenarios of sea level rise uncertainty. As it can be seen from the table, for both dynamic pathways the project value increases as $\sigma$ increases. Moreover, in every considered scenario, the decision to invest in the barrier and dike first, and flood-proofing measures afterwards, provides the highest project benefit.
\begin{table}[h]
    \centering
    \caption{Investment sequences under different discount rate scenarios. We report (in billion USD) the NPV, the total value with optionality and the difference between the two values. For each scenario the project value of the optimal adaptation pathway is indicated in bold letters.}
    \label{sensitivity_portfolio_de_nyc_discount}
    \begin{tabular}{lrrrrrrrrr}
         \hline
         &\multicolumn{9}{c}{Flood-proofing then Barrier and Dike}\\
         \cmidrule{2-10}
         &r=2\%&&r=3\%&&r=4\%&&r=5\%&&r=6\%\\
         \cmidrule{2-2}\cmidrule{4-4}\cmidrule{6-6}\cmidrule{8-8}\cmidrule{10-10}
         NPV & 327.99 &  & 53.99 &  & 8.42 &  & -3.17 &  & -7.27\\
         Total value with optionality & 327.99 &  & 53.99 &  & 10.38 &  & 2.71 &  & \bf{0.85}\\
         Difference & 0.00 &  & 0.00 &  & 1.97 &  & 5.87 &  & 8.11 \\
         \hline
         &\multicolumn{9}{c}{Barrier and Dike then flood-proofing}\\
         \cmidrule{2-10}
         &r=2\%&&r=3\%&&r=4\%&&r=5\%&&r=6\%\\
         \cmidrule{2-2}\cmidrule{4-4}\cmidrule{6-6}\cmidrule{8-8}\cmidrule{10-10}
         NPV & 327.99 &  & 53.99 &  & 8.42 &  & -3.17 &  & -7.27 \\
         Total value with optionality & \bf{338.45} &  & \bf{55.84} &  & \bf{10.69} &  & \bf{2.73} &  & 0.84 \\
         Difference & 10.47 &  & 1.85 &  & 2.27 &  & 5.90 &  & 8.11  \\
         \hline
    \end{tabular}

\end{table}

\begin{table}[h]
    \centering
    \caption{Investment sequences under different scenarios for mean sea level rise. We report (in billion USD) the NPV, the total value with optionality and the difference between the two values. For each scenario the project value of the optimal adaptation pathway is indicated in bold letters.}
    \label{sensitivity_portfolio_de_nyc_slr}
    \begin{tabular}{lrrrrrrrrr}
         \hline
         &\multicolumn{9}{c}{Flood-proofing then Barrier and Dike}\\
         \cmidrule{2-10}
         &$\mu$=0&&$\mu$=3&&$\mu$=6&&$\mu$=9&&$\mu$=12\\
         \cmidrule{2-2}\cmidrule{4-4}\cmidrule{6-6}\cmidrule{8-8}\cmidrule{10-10}
         NPV &  -7.89 &  & -3.50 &  & 8.42 &  & 29.76 &  & 55.48\\
         Total value with optionality & \bf{0.20} &  & \bf{1.95} &  & 10.38 &  & 29.77 &  & 55.48\\
         Difference & 8.09 &  & 5.44 &  & 1.97 &  & 0.02 &  & 0.00\\
         \hline
        &\multicolumn{9}{c}{Barrier and Dike then flood-proofing}\\
        \cmidrule{2-10}
         &$\mu$=0&&$\mu$=3&&$\mu$=6&&$\mu$=9&&$\mu$=12\\
         \cmidrule{2-2}\cmidrule{4-4}\cmidrule{6-6}\cmidrule{8-8}\cmidrule{10-10}
        NPV &  -7.89 &  & -3.50 &  & 8.42 &  & 29.76 &  & 55.49  \\
        Total value with optionality & 0.15 &  & 1.92 &  & \bf{10.69} &  & \bf{31.02} &  & \bf{57.52} \\
        Difference & 8.04 &  & 5.41 &  & 2.27 &  & 1.26 &  & 2.04 \\
        \hline
    \end{tabular}
\end{table}

\begin{table}[h]
    \centering
    \caption{Investment sequences under different levels of sea level rise uncertainty. We report (in billion USD) the NPV, the total value with optionality and the difference between the two values. For each scenario the project value of the optimal adaptation pathway is indicated in bold letters.}
    \label{sensitivity_portfolio_de_nyc_sigmas}
    \begin{tabular}{lrrrrrrrrrrrrrrrrrrr}
        \hline
        &\multicolumn{9}{c}{Flood-proofing then Barrier and Dike}\\
        \cmidrule{2-10}
        &$\sigma$=7&&$\sigma$=15&&$\sigma$=25&&$\sigma$=30&&$\sigma$=45\\
        \cmidrule{2-2}\cmidrule{4-4}\cmidrule{6-6}\cmidrule{8-8}\cmidrule{10-10}
        NPV &  4.00 &  & 5.39 &  & 8.42 &  & 10.40 &  & 17.71 &\\
        Total value with optionality  & 6.27 &  & 7.57 &  & 10.38 &  & 12.22 &  & 19.05 \\
        Difference  & 2.27 &  & 2.18 &  & 1.97 &  & 1.82 &  & 1.35\\
        \hline
        &\multicolumn{9}{c}{Barrier Dike then flood-proofing}\\
        \cmidrule{2-10}
        &$\sigma$=7&&$\sigma$=15&&$\sigma$=25&&$\sigma$=30&&$\sigma$=45\\
        \cmidrule{2-2}\cmidrule{4-4}\cmidrule{6-6}\cmidrule{8-8}\cmidrule{10-10}
        NPV & 4.00 &  & 5.39 &  & 8.42 &  & 10.40 &  & 17.71 \\
        Total value with optionality   & \textbf{6.42} &  & \textbf{7.77} &  & \textbf{10.69} &  & \textbf{12.59} &  & \textbf{19.68} \\
        Difference & 2.42 &  & 2.38 &  & 2.27 &  & 2.19 &  & 1.97 \\
        \hline
    \end{tabular}
\end{table}

\subsection{Discussion}
 The numerical results of the previous sections demonstrate the  potential of our approach, which combines real option analysis and extreme value theory, to assist decision makers in determining optimal climate adaptation pathways.  When evaluated separately, both of the considered adaptation policies provide a positive return on investment, suggesting that both policies provide enough benefit to be implemented by decision makers. As both policies can provide benefits to the city of New York, some jeopardy can arise on which policy should be implemented, whether a portfolio of strategies might be the better option, or if a sequence of investments should be considered. Table \ref{analysis_nyc_portfolio_de} analyzes the optimal investment sequence in the baseline scenario. The results suggest that there is a quantifiable difference in the benefits produced by the two competing sequences (flood-proofing and then barrier and dike vs barrier and dike and then flood-proofing), and therefore the order in which the adaptation measures are implemented matters. \textcolor{black}{ Moreover, it is not optimal to invest right away, but to wait initially and then invest in the barrier and dike project that provides a more substantial protection. Our results also suggest that under our baseline scenario once the barrier and dike project has been invested, there is no additional benefit of investing into the flood-proofing project at a later stage. An alternative pathway, providing a slightly lower benefit is to invest into the flood-proofing project immediately and then into the barrier and dike project at a later stage.
} This aspect was somewhat overlooked in \cite{aerts2014evaluating} where flood-proofing measures and barrier and dike were considered jointly as a hybrid strategy rather than as a sequence. Nonetheless, we have obtained similar results on the optimal investment timing, indicating that deferring the investment for more invasive projects, such as the barrier and dike, might turn out to be more beneficial than investing in it right away.

 Sensitivity analysis in Tables \ref{sensitivity_portfolio_de_nyc_discount}, \ref{sensitivity_portfolio_de_nyc_slr}, and \ref{sensitivity_portfolio_de_nyc_sigmas} illustrates that the 
 obtained results for our methodology are dependent on the chosen parameter values for key input variables. 
 It can be seen from the tables that the project values are more sensitive to interest rate and sea level rise than to sea level rise uncertainty. In particular, during a low interest rate regime, it is more beneficial to invest first in the more invasive and expensive project and then into the less expensive soft measures. In scenarios with high interest rates (higher than 6\%), investing in the soft measures first is preferred, indicating that when the opportunity cost is high, decision makers might prefer soft measures over hard measures in the short run. However, high interest rates should be used with caution when evaluating climate adaptation policies with a very long time horizon, since  future monetary benefits can be undervalued too harshly \cite[see,][]{aerts2014evaluating}. Similar considerations can be made with respect to the sensitivity analysis in sea level rise. 
\textcolor{black}{While our findings for the optimal adaptation pathway remain unchanged for most scenarios, for very low levels of sea level rise, the order of investment into the projects for an optimal adaptation pathway may be changed.} In optimistic scenarios with no climate change impact ($\mu = 0$) and low climate change impact ($\mu=3)$, the optimal investment strategy is to adopt soft measures first and then implement hard measures. However, with expectations of more substantial sea level rise, the hard measures appear to be necessary.
\section{Conclusions}
\label{conc}
Flood risk and, more generally, the risk of natural hazards as a result of climatic change can have catastrophic consequences for our cities, infrastructure and societies. Thus, reducing the impact of climate change can be considered as one of the key challenges of our time faced by decision and policy makers. The concept of adaptation pathways has proven to be a valuable tool in taking decisions on climate adaptation policies, in developing flexible strategies capable to adapt to new information or to changes in the underlying conditions. Nevertheless, so far adaptation pathways are rarely implemented based on an underlying thorough quantitative analysis, possibly making the selected pathway and sequence of projects strongly subjective. In this study we propose a methodology based on extreme value theory and real option analysis to bridge the gap between adaptation pathways, optimal timing and the quantitative analysis for adaptation projects. Our methodology is flexible enough to study optimal adaptation pathways comprising both soft and hard adaptation measures. The approach also allows to factor in growth in the value of the exposure and uncertainty about the impacts of climate change.

We illustrate our methodology with a  case study, aiming to identify the combination of soft measures, such as flood-proofing, and hard infrastructure, such as the construction of a barrier and dike to best protect the area of NYC.

We also performed a sensitivity analysis to assess the robustness of our results with respect to key variables such as the discount rate, the mean level  of sea level rise, and uncertainty about sea level rise. 

Our proposed methodology based on real option analysis for optimal climate adaptation policies can be very helpful for decision and policy makers tasked with the selection of optimal investment sequences. From a policy implication perspective, our methodology can assist decision makers in selecting the optimal investment sequence and the timing for the chosen combination of climate adaptation technologies. Our results suggest that, depending on the current sea level and the characteristics of the competing adaptation projects, decision makers may choose to defer investment in hard measures and disruptive adaptation projects such as the construction of a barrier and dike system, and wait for climate uncertainty to unfold. However, only for a few of the scenarios considered, the adoption of soft measures seems to provide an overall  adequate level of protection from sea level rise in our case study. \textcolor{black}{ Our study provides an important modeling framework to determine whether to invest in low cost adaptation measures first and preserve investment flexibility or to commit to more effective but also more costly adaptation measures first. As illustrated in our numerical application, the answer is not straightforwards, emphasizing the need for a quantitative examination of the solution.}

The real option framework presented in this manuscript can be easily extended to allow for more than just two climate adaptation projects, and can be applied to other locations than NYC. Nevertheless, a few notes of caution need to be mentioned. Firstly, real option analysis requires the estimation of several key parameters, such as the expected rate of sea level rise and the impact of climate indices, which may require relatively long time series. While significant effort has been made to collect and collate data on sea levels, high quality data may be lacking in other areas or less developed countries. Secondly,  our methodology is based on extreme value theory and relies on data on extreme weather events, which are often scarce. While the block maxima approach is a valid and widely accepted method to deal with such data, it is important for researchers and policy makers to consider this limitation. Thirdly, our approach requires reliable estimates of the flood loss curve, which are usually obtained from comprehensive and complex studies. In addition, real option analysis could be quite computationally intensive for more complex portfolios of adaptation projects.

It should be noted that in our proposed methodology, the transmission of climate uncertainty is achieved via the mean sea level. However, future studies may consider extending the model to a two-factor approach where climate uncertainty affects both the mean sea level and storm surge volatility.

\section*{Acknowledgements}

This research was supported by the Society of Actuaries (SOA), under project ``Flood risk management and adaptation under sea-level rise uncertainty'', and Data 61 CSIRO Australia, under project ``Optimal decision making for risk mitigation of natural hazards using a real options approach''. We would like to acknowledge valuable discussions with the SOA project oversight group (Rob Montgomery, Josh Rekula, Remi Villeneuve, Matthew Self, Tamara Wilt, Cindy Bruyere, Priya Rohatgi, Sam Gutterman, Bronwyn Claire), CSIRO research partners Simon Dunstall and Mahesh Prakash, and Babson College Professor Michael Goldstein.

\section*{Compliance with Ethical Standards}

\textbf{Conflict of interest:} The authors declare that there is no conflict of interest.

\textbf{Research involving human participants and/or animals:} This article does not contain any studies with human
participants or animals performed by any of the authors.

\textbf{Informed consent:} This research was supported by funding through the Society of Actuaries (project ``Flood risk management and adaptation under sea-level rise uncertainty'') and Data 61 CSIRO Australia (project ``Optimal decision making for risk mitigation of natural hazards using real options approach'').\\

\section*{Appendix}
\subsection*{Appendix A. Insurance Premium Calculation}
\label{appendix:a}
\textcolor{black}{The insurance premium calculation exercise is carried out with the intent to assign a monetary value to the benefit derived from investing into climate adaptation policy. Nevertheless, considering a city council willing to insure the entirety of a city or district might be impractical. The introduction of a top cover limit, where losses due to floods are insured only up to a pre-specified amount, might provide a more accurate representation of the real world insurance perspective. In this case, one could replace Equation \ref{premium} with the premium of a insurance policy with a top cover limit. For instance, one could consider a decision maker willing to insure losses due to floodwater damage, for one year, up to a pre-specified amount $l$. The corresponding insurance premium would then be:
\begin{align}
	\pi(u,u^*,t,\alpha;L)  &= (1+\delta)e^{\gamma t} \left[ \int_{u}^{\infty}L(M)dH(M) - \int_{m}^{\infty}L(M)dH(M) \right],\notag\\
	&  = \pi(u,t,\alpha;L) - \pi(m,t,\alpha;L),\notag
\end{align}
where $L(m) = l$.
}
\subsection*{Appendix B. Flood loss curve}
\label{appendix:b}
\textcolor{black}{The quadratic form of the loss curve in Equation \ref{quad1} is not necessary for our methodology, and alternative parameterizations of the flood loss curve may be considered as well. One only needs to assure that the integral in Equation \ref{premium} is well defined. For example consider the following simple function:
\begin{equation*}
    L(M)  = \begin{cases}
			0, & \text{if } M<u^*\\
                L_1, & \text{if } u^*\leq M< M_1\\
                L_2, & \text{if } M_1\leq M< M_2\\
                L_3, & \text{if } M\geq M_2\\
		 \end{cases},
\end{equation*}
where the monetary losses corresponding to floodwater levels are discretized into 4 categories $0,L_1,L_2,L_3$ and with 3 distinct levels for each damage $u^*, M_1, M_2$.}

\bibliography{sn-bibliography}


\begin{thebibliography}{71}
\ifx \bisbn   \undefined \def \bisbn  #1{ISBN #1}\fi
\ifx \binits  \undefined \def \binits#1{#1}\fi
\ifx \bauthor  \undefined \def \bauthor#1{#1}\fi
\ifx \batitle  \undefined \def \batitle#1{#1}\fi
\ifx \bjtitle  \undefined \def \bjtitle#1{#1}\fi
\ifx \bvolume  \undefined \def \bvolume#1{\textbf{#1}}\fi
\ifx \byear  \undefined \def \byear#1{#1}\fi
\ifx \bissue  \undefined \def \bissue#1{#1}\fi
\ifx \bfpage  \undefined \def \bfpage#1{#1}\fi
\ifx \blpage  \undefined \def \blpage #1{#1}\fi
\ifx \burl  \undefined \def \burl#1{\textsf{#1}}\fi
\ifx \doiurl  \undefined \def \doiurl#1{\url{https://doi.org/#1}}\fi
\ifx \betal  \undefined \def \betal{\textit{et al.}}\fi
\ifx \binstitute  \undefined \def \binstitute#1{#1}\fi
\ifx \binstitutionaled  \undefined \def \binstitutionaled#1{#1}\fi
\ifx \bctitle  \undefined \def \bctitle#1{#1}\fi
\ifx \beditor  \undefined \def \beditor#1{#1}\fi
\ifx \bpublisher  \undefined \def \bpublisher#1{#1}\fi
\ifx \bbtitle  \undefined \def \bbtitle#1{#1}\fi
\ifx \bedition  \undefined \def \bedition#1{#1}\fi
\ifx \bseriesno  \undefined \def \bseriesno#1{#1}\fi
\ifx \blocation  \undefined \def \blocation#1{#1}\fi
\ifx \bsertitle  \undefined \def \bsertitle#1{#1}\fi
\ifx \bsnm \undefined \def \bsnm#1{#1}\fi
\ifx \bsuffix \undefined \def \bsuffix#1{#1}\fi
\ifx \bparticle \undefined \def \bparticle#1{#1}\fi
\ifx \barticle \undefined \def \barticle#1{#1}\fi
\bibcommenthead
\ifx \bconfdate \undefined \def \bconfdate #1{#1}\fi
\ifx \botherref \undefined \def \botherref #1{#1}\fi
\ifx \url \undefined \def \url#1{\textsf{#1}}\fi
\ifx \bchapter \undefined \def \bchapter#1{#1}\fi
\ifx \bbook \undefined \def \bbook#1{#1}\fi
\ifx \bcomment \undefined \def \bcomment#1{#1}\fi
\ifx \oauthor \undefined \def \oauthor#1{#1}\fi
\ifx \citeauthoryear \undefined \def \citeauthoryear#1{#1}\fi
\ifx \endbibitem  \undefined \def \endbibitem {}\fi
\ifx \bconflocation  \undefined \def \bconflocation#1{#1}\fi
\ifx \arxivurl  \undefined \def \arxivurl#1{\textsf{#1}}\fi
\csname PreBibitemsHook\endcsname

\bibitem{portner2022climate}
\begin{botherref}
\oauthor{\bsnm{P{\"o}rtner}, \binits{H.-O.}},
\oauthor{\bsnm{Roberts}, \binits{D.C.}},
\oauthor{\bsnm{Adams}, \binits{H.}},
\oauthor{\bsnm{Adler}, \binits{C.}},
\oauthor{\bsnm{Aldunce}, \binits{P.}},
\oauthor{\bsnm{Ali}, \binits{E.}},
\oauthor{\bsnm{Begum}, \binits{R.A.}},
\oauthor{\bsnm{Betts}, \binits{R.}},
\oauthor{\bsnm{Kerr}, \binits{R.B.}},
\oauthor{\bsnm{Biesbroek}, \binits{R.}}, et al.:
Climate change 2022: Impacts, adaptation and vulnerability.
IPCC Sixth Assessment Report
(2022)
\end{botherref}
\endbibitem

\bibitem{wang2015adaptation}
\begin{barticle}
\bauthor{\bsnm{Wang}, \binits{C.-H.}},
\bauthor{\bsnm{Khoo}, \binits{Y.B.}},
\bauthor{\bsnm{Wang}, \binits{X.}}:
\batitle{Adaptation benefits and costs of raising coastal buildings under
  storm-tide inundation in {South East Queensland, Australia}}.
\bjtitle{Climatic Change}
\bvolume{132}(\bissue{4}),
\bfpage{545}--\blpage{558}
(\byear{2015})
\end{barticle}
\endbibitem

\bibitem{rade2017}
\begin{barticle}
\bauthor{\bsnm{Musulin}, \binits{R.}}:
\batitle{The reality of flood insurance repayment}.
\bjtitle{Casualty Quarterly}
\bvolume{1}(\bissue{3}),
\bfpage{1}--\blpage{3}
(\byear{2017})
\end{barticle}
\endbibitem

\bibitem{song2019}
\begin{barticle}
\bauthor{\bsnm{Song}, \binits{M.}},
\bauthor{\bsnm{Du}, \binits{Q.}}:
\batitle{{Analysis and exploration of damage-reduction measures for flood
  disasters in China}}.
\bjtitle{Annals of Operations Research}
\bvolume{283}(\bissue{1}),
\bfpage{795}--\blpage{810}
(\byear{2019})
\end{barticle}
\endbibitem

\bibitem{al2020}
\begin{botherref}
\oauthor{\bsnm{Al~Qundus}, \binits{J.}},
\oauthor{\bsnm{Dabbour}, \binits{K.}},
\oauthor{\bsnm{Gupta}, \binits{S.}},
\oauthor{\bsnm{Meissonier}, \binits{R.}},
\oauthor{\bsnm{Paschke}, \binits{A.}}:
{Wireless sensor network for AI-based flood disaster detection}.
Annals of Operations Research,
1--23
(2020)
\end{botherref}
\endbibitem

\bibitem{fazey2016}
\begin{barticle}
\bauthor{\bsnm{Fazey}, \binits{I.}},
\bauthor{\bsnm{Wise}, \binits{R.M.}},
\bauthor{\bsnm{Lyon}, \binits{C.}},
\bauthor{\bsnm{C{\^a}mpeanu}, \binits{C.}},
\bauthor{\bsnm{Moug}, \binits{P.}},
\bauthor{\bsnm{Davies}, \binits{T.E.}}:
\batitle{Past and future adaptation pathways}.
\bjtitle{Climate and Development}
\bvolume{8}(\bissue{1}),
\bfpage{26}--\blpage{44}
(\byear{2016})
\end{barticle}
\endbibitem

\bibitem{buurman2016}
\begin{barticle}
\bauthor{\bsnm{Buurman}, \binits{J.}},
\bauthor{\bsnm{Babovic}, \binits{V.}}:
\batitle{Adaptation pathways and real options analysis: An approach to deep
  uncertainty in climate change adaptation policies}.
\bjtitle{Policy and Society}
\bvolume{35}(\bissue{2}),
\bfpage{137}--\blpage{150}
(\byear{2016})
\end{barticle}
\endbibitem

\bibitem{werners2021}
\begin{barticle}
\bauthor{\bsnm{Werners}, \binits{S.E.}},
\bauthor{\bsnm{Wise}, \binits{R.M.}},
\bauthor{\bsnm{Butler}, \binits{J.R.}},
\bauthor{\bsnm{Totin}, \binits{E.}},
\bauthor{\bsnm{Vincent}, \binits{K.}}:
\batitle{Adaptation pathways: A review of approaches and a learning framework}.
\bjtitle{Environmental Science \& Policy}
\bvolume{116},
\bfpage{266}--\blpage{275}
(\byear{2021})
\end{barticle}
\endbibitem

\bibitem{cradock2020}
\begin{barticle}
\bauthor{\bsnm{Cradock-Henry}, \binits{N.A.}},
\bauthor{\bsnm{Blackett}, \binits{P.}},
\bauthor{\bsnm{Hall}, \binits{M.}},
\bauthor{\bsnm{Johnstone}, \binits{P.}},
\bauthor{\bsnm{Teixeira}, \binits{E.}},
\bauthor{\bsnm{Wreford}, \binits{A.}}:
\batitle{Climate adaptation pathways for agriculture: insights from a
  participatory process}.
\bjtitle{Environmental Science \& Policy}
\bvolume{107},
\bfpage{66}--\blpage{79}
(\byear{2020})
\end{barticle}
\endbibitem

\bibitem{haasnoot2013}
\begin{barticle}
\bauthor{\bsnm{Haasnoot}, \binits{M.}},
\bauthor{\bsnm{Kwakkel}, \binits{J.H.}},
\bauthor{\bsnm{Walker}, \binits{W.E.}},
\bauthor{\bsnm{Ter~Maat}, \binits{J.}}:
\batitle{Dynamic adaptive policy pathways: A method for crafting robust
  decisions for a deeply uncertain world}.
\bjtitle{Global Environmental Change}
\bvolume{23}(\bissue{2}),
\bfpage{485}--\blpage{498}
(\byear{2013})
\end{barticle}
\endbibitem

\bibitem{ranger2013}
\begin{barticle}
\bauthor{\bsnm{Ranger}, \binits{N.}},
\bauthor{\bsnm{Reeder}, \binits{T.}},
\bauthor{\bsnm{Lowe}, \binits{J.}}:
\batitle{{Addressing ‘deep’uncertainty over long-term climate in major
  infrastructure projects: four innovations of the Thames Estuary 2100
  Project}}.
\bjtitle{EURO Journal on Decision Processes}
\bvolume{1}(\bissue{3-4}),
\bfpage{233}--\blpage{262}
(\byear{2013})
\end{barticle}
\endbibitem

\bibitem{kim2019}
\begin{barticle}
\bauthor{\bsnm{Kim}, \binits{M.-J.}},
\bauthor{\bsnm{Nicholls}, \binits{R.J.}},
\bauthor{\bsnm{Preston}, \binits{J.M.}},
\bauthor{\bparticle{de} \bsnm{Almeida}, \binits{G.A.}}:
\batitle{An assessment of the optimum timing of coastal flood adaptation given
  sea-level rise using real options analysis}.
\bjtitle{Journal of Flood Risk Management}
\bvolume{12}(\bissue{S2}),
\bfpage{12494}
(\byear{2019})
\end{barticle}
\endbibitem

\bibitem{dobes2010}
\begin{botherref}
\oauthor{\bsnm{Dobes}, \binits{L.}}:
{Notes on applying ‘real options’ to climate change adaptation measures,
  with examples from Vietnam}.
Crawford School Research Paper, Crawford School of Economics and
  Government--Centre for Climate Economics \& Policy (CCEP) Working Paper
\textbf{7}
(2010)
\end{botherref}
\endbibitem

\bibitem{nicholls2014}
\begin{barticle}
\bauthor{\bsnm{Nicholls}, \binits{R.J.}},
\bauthor{\bsnm{Hanson}, \binits{S.E.}},
\bauthor{\bsnm{Lowe}, \binits{J.A.}},
\bauthor{\bsnm{Warrick}, \binits{R.A.}},
\bauthor{\bsnm{Lu}, \binits{X.}},
\bauthor{\bsnm{Long}, \binits{A.J.}}:
\batitle{Sea-level scenarios for evaluating coastal impacts}.
\bjtitle{Wiley Interdisciplinary Reviews: Climate Change}
\bvolume{5}(\bissue{1}),
\bfpage{129}--\blpage{150}
(\byear{2014})
\end{barticle}
\endbibitem

\bibitem{downing2012}
\begin{barticle}
\bauthor{\bsnm{Downing}, \binits{T.E.}}:
\batitle{Views of the frontiers in climate change adaptation economics}.
\bjtitle{Wiley Interdisciplinary Reviews: Climate Change}
\bvolume{3}(\bissue{2}),
\bfpage{161}--\blpage{170}
(\byear{2012})
\end{barticle}
\endbibitem

\bibitem{haasnoot2012}
\begin{barticle}
\bauthor{\bsnm{Haasnoot}, \binits{M.}},
\bauthor{\bsnm{Middelkoop}, \binits{H.}},
\bauthor{\bsnm{Offermans}, \binits{A.}},
\bauthor{\bsnm{Van~Beek}, \binits{E.}},
\bauthor{\bsnm{Van~Deursen}, \binits{W.P.}}:
\batitle{Exploring pathways for sustainable water management in river deltas in
  a changing environment}.
\bjtitle{Climatic Change}
\bvolume{115}(\bissue{3}),
\bfpage{795}--\blpage{819}
(\byear{2012})
\end{barticle}
\endbibitem

\bibitem{wise2014}
\begin{barticle}
\bauthor{\bsnm{Wise}, \binits{R.M.}},
\bauthor{\bsnm{Fazey}, \binits{I.}},
\bauthor{\bsnm{Smith}, \binits{M.S.}},
\bauthor{\bsnm{Park}, \binits{S.E.}},
\bauthor{\bsnm{Eakin}, \binits{H.}},
\bauthor{\bsnm{Van~Garderen}, \binits{E.A.}},
\bauthor{\bsnm{Campbell}, \binits{B.}}:
\batitle{Reconceptualising adaptation to climate change as part of pathways of
  change and response}.
\bjtitle{Global environmental change}
\bvolume{28},
\bfpage{325}--\blpage{336}
(\byear{2014})
\end{barticle}
\endbibitem

\bibitem{brandao2005}
\begin{barticle}
\bauthor{\bsnm{Brandao}, \binits{L.E.}},
\bauthor{\bsnm{Dyer}, \binits{J.S.}}:
\batitle{Decision analysis and real options: A discrete time approach to real
  option valuation}.
\bjtitle{Annals of Operations Research}
\bvolume{135}(\bissue{1}),
\bfpage{21}--\blpage{39}
(\byear{2005})
\end{barticle}
\endbibitem

\bibitem{woodward2011}
\begin{barticle}
\bauthor{\bsnm{Woodward}, \binits{M.}},
\bauthor{\bsnm{Gouldby}, \binits{B.}},
\bauthor{\bsnm{Kapelan}, \binits{Z.}},
\bauthor{\bsnm{Khu}, \binits{S.-T.}},
\bauthor{\bsnm{Townend}, \binits{I.}}:
\batitle{Real options in flood risk management decision making}.
\bjtitle{Journal of Flood Risk Management}
\bvolume{4}(\bissue{4}),
\bfpage{339}--\blpage{349}
(\byear{2011})
\end{barticle}
\endbibitem

\bibitem{woodward2014}
\begin{barticle}
\bauthor{\bsnm{Woodward}, \binits{M.}},
\bauthor{\bsnm{Kapelan}, \binits{Z.}},
\bauthor{\bsnm{Gouldby}, \binits{B.}}:
\batitle{Adaptive flood risk management under climate change uncertainty using
  real options and optimization}.
\bjtitle{Risk Analysis}
\bvolume{34}(\bissue{1}),
\bfpage{75}--\blpage{92}
(\byear{2014})
\end{barticle}
\endbibitem

\bibitem{wreford2020}
\begin{barticle}
\bauthor{\bsnm{Wreford}, \binits{A.}},
\bauthor{\bsnm{Dittrich}, \binits{R.}},
\bauthor{\bparticle{van~der} \bsnm{Pol}, \binits{T.D.}}:
\batitle{The added value of real options analysis for climate change
  adaptation}.
\bjtitle{Wiley Interdisciplinary Reviews: Climate Change}
\bvolume{11}(\bissue{3}),
\bfpage{642}
(\byear{2020})
\end{barticle}
\endbibitem

\bibitem{park2014}
\begin{barticle}
\bauthor{\bsnm{Park}, \binits{T.}},
\bauthor{\bsnm{Kim}, \binits{C.}},
\bauthor{\bsnm{Kim}, \binits{H.}}:
\batitle{Valuation of drainage infrastructure improvement under climate change
  using real options}.
\bjtitle{Water Resources Management}
\bvolume{28}(\bissue{2}),
\bfpage{445}--\blpage{457}
(\byear{2014})
\end{barticle}
\endbibitem

\bibitem{chan2016dynamic}
\begin{barticle}
\bauthor{\bsnm{Chan}, \binits{R.}},
\bauthor{\bsnm{Durango-Cohen}, \binits{P.L.}},
\bauthor{\bsnm{Schofer}, \binits{J.L.}}:
\batitle{Dynamic learning process for selecting storm protection investments}.
\bjtitle{Transportation Research Record}
\bvolume{2599}(\bissue{1}),
\bfpage{1}--\blpage{8}
(\byear{2016})
\end{barticle}
\endbibitem

\bibitem{oh2018investment}
\begin{barticle}
\bauthor{\bsnm{Oh}, \binits{S.}},
\bauthor{\bsnm{Kim}, \binits{K.}},
\bauthor{\bsnm{Kim}, \binits{H.}}:
\batitle{{Investment decision for coastal urban development projects
  considering the impact of climate change: Case study of the Great Garuda
  Project in Indonesia}}.
\bjtitle{Journal of Cleaner Production}
\bvolume{178},
\bfpage{507}--\blpage{514}
(\byear{2018})
\end{barticle}
\endbibitem

\bibitem{kim2018}
\begin{barticle}
\bauthor{\bsnm{Kim}, \binits{K.}},
\bauthor{\bsnm{Kim}, \binits{J.-S.}}:
\batitle{{Economic assessment of flood control facilities under climate
  uncertainty: A case of Nakdong River, South Korea}}.
\bjtitle{Sustainability}
\bvolume{10}(\bissue{2}),
\bfpage{308}
(\byear{2018})
\end{barticle}
\endbibitem

\bibitem{regan2017}
\begin{barticle}
\bauthor{\bsnm{Regan}, \binits{C.M.}},
\bauthor{\bsnm{Connor}, \binits{J.D.}},
\bauthor{\bsnm{Segaran}, \binits{R.R.}},
\bauthor{\bsnm{Meyer}, \binits{W.S.}},
\bauthor{\bsnm{Bryan}, \binits{B.A.}},
\bauthor{\bsnm{Ostendorf}, \binits{B.}}:
\batitle{Climate change and the economics of biomass energy feedstocks in
  semi-arid agricultural landscapes: A spatially explicit real options
  analysis}.
\bjtitle{Journal of Environmental Management}
\bvolume{192},
\bfpage{171}--\blpage{183}
(\byear{2017})
\end{barticle}
\endbibitem

\bibitem{schiel2019real}
\begin{barticle}
\bauthor{\bsnm{Schiel}, \binits{C.}},
\bauthor{\bsnm{Gl{\"o}ser-Chahoud}, \binits{S.}},
\bauthor{\bsnm{Schultmann}, \binits{F.}}:
\batitle{A real option application for emission control measures}.
\bjtitle{Journal of Business Economics}
\bvolume{89}(\bissue{3}),
\bfpage{291}--\blpage{325}
(\byear{2019})
\end{barticle}
\endbibitem

\bibitem{ginbo2021}
\begin{barticle}
\bauthor{\bsnm{Ginbo}, \binits{T.}},
\bauthor{\bsnm{Di~Corato}, \binits{L.}},
\bauthor{\bsnm{Hoffmann}, \binits{R.}}:
\batitle{Investing in climate change adaptation and mitigation: A
  methodological review of real-options studies}.
\bjtitle{Ambio}
\bvolume{50}(\bissue{1}),
\bfpage{229}--\blpage{241}
(\byear{2021})
\end{barticle}
\endbibitem

\bibitem{gersonius2011failure}
\begin{barticle}
\bauthor{\bsnm{Gersonius}, \binits{B.}},
\bauthor{\bsnm{Morselt}, \binits{T.}},
\bauthor{\bsnm{Van~Nieuwenhuijzen}, \binits{L.}},
\bauthor{\bsnm{Ashley}, \binits{R.}},
\bauthor{\bsnm{Zevenbergen}, \binits{C.}}:
\batitle{How the failure to account for flexibility in the economic analysis of
  flood risk and coastal management strategies can result in maladaptive
  decisions}.
\bjtitle{Journal of Waterway, Port, Coastal, and Ocean Engineering}
\bvolume{138}(\bissue{5}),
\bfpage{386}--\blpage{393}
(\byear{2011})
\end{barticle}
\endbibitem

\bibitem{munoz2011}
\begin{barticle}
\bauthor{\bsnm{Mu{\~n}oz}, \binits{J.I.}},
\bauthor{\bsnm{Contreras}, \binits{J.}},
\bauthor{\bsnm{Caama{\~n}o}, \binits{J.}},
\bauthor{\bsnm{Correia}, \binits{P.}}:
\batitle{A decision-making tool for project investments based on real options:
  the case of wind power generation}.
\bjtitle{Annals of Operations Research}
\bvolume{186}(\bissue{1}),
\bfpage{465}--\blpage{490}
(\byear{2011})
\end{barticle}
\endbibitem

\bibitem{chesney2017}
\begin{barticle}
\bauthor{\bsnm{Chesney}, \binits{M.}},
\bauthor{\bsnm{Lasserre}, \binits{P.}},
\bauthor{\bsnm{Troja}, \binits{B.}}:
\batitle{Mitigating global warming: A real options approach}.
\bjtitle{Annals of operations research}
\bvolume{255}(\bissue{1}),
\bfpage{465}--\blpage{506}
(\byear{2017})
\end{barticle}
\endbibitem

\bibitem{mac2020}
\begin{barticle}
\bauthor{\bsnm{Mac~Cawley}, \binits{A.}},
\bauthor{\bsnm{Cubillos}, \binits{M.}},
\bauthor{\bsnm{Pascual}, \binits{R.}}:
\batitle{A real options approach for joint overhaul and replacement strategies
  with mean reverting prices}.
\bjtitle{Annals of Operations Research}
\bvolume{286}(\bissue{1}),
\bfpage{303}--\blpage{324}
(\byear{2020})
\end{barticle}
\endbibitem

\bibitem{truong2016}
\begin{botherref}
\oauthor{\bsnm{Truong}, \binits{C.}},
\oauthor{\bsnm{Tr\"uck}, \binits{S.}}:
It's not now or never: Implications of investment timing and risk aversion on
  climate adaptation to extreme events.
European Journal of Operational Research,
(2016).
\doiurl{10.1016/j.ejor.2016.01.044}
\end{botherref}
\endbibitem

\bibitem{truong2017managing}
\begin{botherref}
\oauthor{\bsnm{Truong}, \binits{C.}},
\oauthor{\bsnm{Tr{\"u}ck}, \binits{S.}},
\oauthor{\bsnm{Mathew}, \binits{S.}}:
Managing risks from climate impacted hazards-the value of investment
  flexibility under uncertainty.
European Journal of Operational Research
(2017)
\end{botherref}
\endbibitem

\bibitem{menendez2010}
\begin{botherref}
\oauthor{\bsnm{Men{\'e}ndez}, \binits{M.}},
\oauthor{\bsnm{Woodworth}, \binits{P.L.}}:
Changes in extreme high water levels based on a quasi-global tide-gauge data
  set.
Journal of Geophysical Research: Oceans
\textbf{115}(C10)
(2010)
\end{botherref}
\endbibitem

\bibitem{lobeto2018}
\begin{barticle}
\bauthor{\bsnm{Lobeto}, \binits{H.}},
\bauthor{\bsnm{Menendez}, \binits{M.}},
\bauthor{\bsnm{Losada}, \binits{I.}}:
\batitle{Toward a methodology for estimating coastal extreme sea levels from
  satellite altimetry}.
\bjtitle{Journal of Geophysical Research: Oceans}
\bvolume{123}(\bissue{11}),
\bfpage{8284}--\blpage{8298}
(\byear{2018})
\end{barticle}
\endbibitem

\bibitem{salas2018}
\begin{barticle}
\bauthor{\bsnm{Salas}, \binits{J.}},
\bauthor{\bsnm{Obeysekera}, \binits{J.}},
\bauthor{\bsnm{Vogel}, \binits{R.}}:
\batitle{Techniques for assessing water infrastructure for nonstationary
  extreme events: A review}.
\bjtitle{Hydrological Sciences Journal}
\bvolume{63}(\bissue{3}),
\bfpage{325}--\blpage{352}
(\byear{2018})
\end{barticle}
\endbibitem

\bibitem{aerts2014evaluating}
\begin{barticle}
\bauthor{\bsnm{Aerts}, \binits{J.C.}},
\bauthor{\bsnm{Botzen}, \binits{W.W.}},
\bauthor{\bsnm{Emanuel}, \binits{K.}},
\bauthor{\bsnm{Lin}, \binits{N.}},
\bauthor{\bsnm{De~Moel}, \binits{H.}},
\bauthor{\bsnm{Michel-Kerjan}, \binits{E.O.}}:
\batitle{Evaluating flood resilience strategies for coastal megacities}.
\bjtitle{Science}
\bvolume{344}(\bissue{6183}),
\bfpage{473}--\blpage{475}
(\byear{2014})
\end{barticle}
\endbibitem

\bibitem{gersonius2013climate}
\begin{barticle}
\bauthor{\bsnm{Gersonius}, \binits{B.}},
\bauthor{\bsnm{Ashley}, \binits{R.}},
\bauthor{\bsnm{Pathirana}, \binits{A.}},
\bauthor{\bsnm{Zevenbergen}, \binits{C.}}:
\batitle{Climate change uncertainty: building flexibility into water and flood
  risk infrastructure}.
\bjtitle{Climatic Change}
\bvolume{116}(\bissue{2}),
\bfpage{411}--\blpage{423}
(\byear{2013})
\end{barticle}
\endbibitem

\bibitem{kim2017using}
\begin{barticle}
\bauthor{\bsnm{Kim}, \binits{K.}},
\bauthor{\bsnm{Ha}, \binits{S.}},
\bauthor{\bsnm{Kim}, \binits{H.}}:
\batitle{Using real options for urban infrastructure adaptation under climate
  change}.
\bjtitle{Journal of Cleaner Production}
\bvolume{143},
\bfpage{40}--\blpage{50}
(\byear{2017})
\end{barticle}
\endbibitem

\bibitem{brown2018}
\begin{barticle}
\bauthor{\bsnm{Brown}, \binits{J.M.}},
\bauthor{\bsnm{Morrissey}, \binits{K.}},
\bauthor{\bsnm{Knight}, \binits{P.}},
\bauthor{\bsnm{Prime}, \binits{T.D.}},
\bauthor{\bsnm{Almeida}, \binits{L.P.}},
\bauthor{\bsnm{Masselink}, \binits{G.}},
\bauthor{\bsnm{Bird}, \binits{C.O.}},
\bauthor{\bsnm{Dodds}, \binits{D.}},
\bauthor{\bsnm{Plater}, \binits{A.J.}}:
\batitle{A coastal vulnerability assessment for planning climate resilient
  infrastructure}.
\bjtitle{Ocean \& Coastal Management}
\bvolume{163},
\bfpage{101}--\blpage{112}
(\byear{2018})
\end{barticle}
\endbibitem

\bibitem{dittrich2019making}
\begin{barticle}
\bauthor{\bsnm{Dittrich}, \binits{R.}},
\bauthor{\bsnm{Butler}, \binits{A.}},
\bauthor{\bsnm{Ball}, \binits{T.}},
\bauthor{\bsnm{Wreford}, \binits{A.}},
\bauthor{\bsnm{Moran}, \binits{D.}}:
\batitle{{Making real options analysis more accessible for climate change
  adaptation. An application to afforestation as a flood management measure in
  the Scottish Borders}}.
\bjtitle{Journal of Environmental Management}
\bvolume{245},
\bfpage{338}--\blpage{347}
(\byear{2019})
\end{barticle}
\endbibitem

\bibitem{salas2014}
\begin{barticle}
\bauthor{\bsnm{Salas}, \binits{J.D.}},
\bauthor{\bsnm{Obeysekera}, \binits{J.}}:
\batitle{Revisiting the concepts of return period and risk for nonstationary
  hydrologic extreme events}.
\bjtitle{Journal of Hydrologic Engineering}
\bvolume{19}(\bissue{3}),
\bfpage{554}--\blpage{568}
(\byear{2014})
\end{barticle}
\endbibitem

\bibitem{hieronymus2020}
\begin{barticle}
\bauthor{\bsnm{Hieronymus}, \binits{M.}},
\bauthor{\bsnm{Kal{\'e}n}, \binits{O.}}:
\batitle{{Sea-level rise projections for Sweden based on the new IPCC special
  report: The ocean and cryosphere in a changing climate}}.
\bjtitle{Ambio}
\bvolume{49}(\bissue{10}),
\bfpage{1587}--\blpage{1600}
(\byear{2020})
\end{barticle}
\endbibitem

\bibitem{muis2018}
\begin{barticle}
\bauthor{\bsnm{Muis}, \binits{S.}},
\bauthor{\bsnm{Haigh}, \binits{I.D.}},
\bauthor{\bsnm{Guimar{\~a}es~Nobre}, \binits{G.}},
\bauthor{\bsnm{Aerts}, \binits{J.C.}},
\bauthor{\bsnm{Ward}, \binits{P.J.}}:
\batitle{Influence of el ni{\~n}o-southern oscillation on global coastal
  flooding}.
\bjtitle{Earth's Future}
\bvolume{6}(\bissue{9}),
\bfpage{1311}--\blpage{1322}
(\byear{2018})
\end{barticle}
\endbibitem

\bibitem{fisher1928limiting}
\begin{bchapter}
\bauthor{\bsnm{Fisher}, \binits{R.A.}},
\bauthor{\bsnm{Tippett}, \binits{L.H.C.}}:
\bctitle{Limiting forms of the frequency distribution of the largest or
  smallest member of a sample}.
In: \bbtitle{Mathematical Proceedings of the Cambridge Philosophical Society},
vol. \bseriesno{24},
pp. \bfpage{180}--\blpage{190}
(\byear{1928}).
\bcomment{Cambridge University Press}
\end{bchapter}
\endbibitem

\bibitem{prettenthaler2010estimation}
\begin{barticle}
\bauthor{\bsnm{Prettenthaler}, \binits{F.}},
\bauthor{\bsnm{Amrusch}, \binits{P.}},
\bauthor{\bsnm{Habsburg-Lothringen}, \binits{C.}}:
\batitle{{Estimation of an absolute flood damage curve based on an Austrian
  case study under a dam breach scenario}}.
\bjtitle{Natural Hazards and Earth System Sciences}
\bvolume{10}(\bissue{4}),
\bfpage{881}--\blpage{894}
(\byear{2010})
\end{barticle}
\endbibitem

\bibitem{merz2004estimation}
\begin{barticle}
\bauthor{\bsnm{Merz}, \binits{B.}},
\bauthor{\bsnm{Kreibich}, \binits{H.}},
\bauthor{\bsnm{Thieken}, \binits{A.}},
\bauthor{\bsnm{Schmidtke}, \binits{R.}}:
\batitle{Estimation uncertainty of direct monetary flood damage to buildings}.
\bjtitle{Natural Hazards and Earth System Sciences}
\bvolume{4}(\bissue{1}),
\bfpage{153}--\blpage{163}
(\byear{2004})
\end{barticle}
\endbibitem

\bibitem{hallegatte2011assessing}
\begin{barticle}
\bauthor{\bsnm{Hallegatte}, \binits{S.}},
\bauthor{\bsnm{Ranger}, \binits{N.}},
\bauthor{\bsnm{Mestre}, \binits{O.}},
\bauthor{\bsnm{Dumas}, \binits{P.}},
\bauthor{\bsnm{Corfee-Morlot}, \binits{J.}},
\bauthor{\bsnm{Herweijer}, \binits{C.}},
\bauthor{\bsnm{Wood}, \binits{R.M.}}:
\batitle{{Assessing climate change impacts, sea level rise and storm surge risk
  in port cities: a case study on Copenhagen}}.
\bjtitle{Climatic Change}
\bvolume{104}(\bissue{1}),
\bfpage{113}--\blpage{137}
(\byear{2011})
\end{barticle}
\endbibitem

\bibitem{buchele2006flood}
\begin{barticle}
\bauthor{\bsnm{B{\"u}chele}, \binits{B.}},
\bauthor{\bsnm{Kreibich}, \binits{H.}},
\bauthor{\bsnm{Kron}, \binits{A.}},
\bauthor{\bsnm{Thieken}, \binits{A.}},
\bauthor{\bsnm{Ihringer}, \binits{J.}},
\bauthor{\bsnm{Oberle}, \binits{P.}},
\bauthor{\bsnm{Merz}, \binits{B.}},
\bauthor{\bsnm{Nestmann}, \binits{F.}}:
\batitle{Flood-risk mapping: contributions towards an enhanced assessment of
  extreme events and associated risks}.
\bjtitle{Natural Hazards and Earth System Sciences}
\bvolume{6}(\bissue{4}),
\bfpage{485}--\blpage{503}
(\byear{2006})
\end{barticle}
\endbibitem

\bibitem{parker2007new}
\begin{barticle}
\bauthor{\bsnm{Parker}, \binits{D.J.}},
\bauthor{\bsnm{Tunstall}, \binits{S.M.}},
\bauthor{\bsnm{McCarthy}, \binits{S.}}:
\batitle{{New insights into the benefits of flood warnings: Results from a
  household survey in England and Wales}}.
\bjtitle{Environmental Hazards}
\bvolume{7}(\bissue{3}),
\bfpage{193}--\blpage{210}
(\byear{2007})
\end{barticle}
\endbibitem

\bibitem{han2020agent}
\begin{barticle}
\bauthor{\bsnm{Han}, \binits{Y.}},
\bauthor{\bsnm{Ash}, \binits{K.}},
\bauthor{\bsnm{Mao}, \binits{L.}},
\bauthor{\bsnm{Peng}, \binits{Z.-R.}}:
\batitle{An agent-based model for community flood adaptation under uncertain
  sea-level rise}.
\bjtitle{Climatic Change}
\bvolume{162}(\bissue{4}),
\bfpage{2257}--\blpage{2276}
(\byear{2020})
\end{barticle}
\endbibitem

\bibitem{merz2010}
\begin{barticle}
\bauthor{\bsnm{Merz}, \binits{B.}},
\bauthor{\bsnm{Kreibich}, \binits{H.}},
\bauthor{\bsnm{Schwarze}, \binits{R.}},
\bauthor{\bsnm{Thieken}, \binits{A.}}:
\batitle{Assessment of economic flood damage}.
\bjtitle{Natural Hazards and Earth System Sciences}
\bvolume{10}(\bissue{8}),
\bfpage{1697}--\blpage{1724}
(\byear{2010})
\end{barticle}
\endbibitem

\bibitem{seifert2010}
\begin{barticle}
\bauthor{\bsnm{Seifert}, \binits{I.}},
\bauthor{\bsnm{Kreibich}, \binits{H.}},
\bauthor{\bsnm{Merz}, \binits{B.}},
\bauthor{\bsnm{Thieken}, \binits{A.H.}}:
\batitle{Application and validation of flemocs--a flood-loss estimation model
  for the commercial sector}.
\bjtitle{Hydrological Sciences Journal--Journal des Sciences Hydrologiques}
\bvolume{55}(\bissue{8}),
\bfpage{1315}--\blpage{1324}
(\byear{2010})
\end{barticle}
\endbibitem

\bibitem{kreibich2010}
\begin{barticle}
\bauthor{\bsnm{Kreibich}, \binits{H.}},
\bauthor{\bsnm{Seifert}, \binits{I.}},
\bauthor{\bsnm{Merz}, \binits{B.}},
\bauthor{\bsnm{Thieken}, \binits{A.H.}}:
\batitle{Development of flemocs--a new model for the estimation of flood losses
  in the commercial sector}.
\bjtitle{Hydrological Sciences Journal--Journal des Sciences Hydrologiques}
\bvolume{55}(\bissue{8}),
\bfpage{1302}--\blpage{1314}
(\byear{2010})
\end{barticle}
\endbibitem

\bibitem{gerl2016}
\begin{barticle}
\bauthor{\bsnm{Gerl}, \binits{T.}},
\bauthor{\bsnm{Kreibich}, \binits{H.}},
\bauthor{\bsnm{Franco}, \binits{G.}},
\bauthor{\bsnm{Marechal}, \binits{D.}},
\bauthor{\bsnm{Schr{\"o}ter}, \binits{K.}}:
\batitle{A review of flood loss models as basis for harmonization and
  benchmarking}.
\bjtitle{PLOS ONE}
\bvolume{11}(\bissue{7}),
\bfpage{1}--\blpage{22}
(\byear{2016})
\end{barticle}
\endbibitem

\bibitem{el2013parameters}
\begin{botherref}
\oauthor{\bsnm{El~Sherpieny}, \binits{E.}},
\oauthor{\bsnm{Assar}, \binits{S.}},
\oauthor{\bsnm{Amer}, \binits{N.}}:
{Parameters Estimation of Generalized Extreme Value Distribution under
  Progressive Type II Censored}.
Asian Journal of Applied Sciences, Vol.2, Issue 2.
(2013)
\end{botherref}
\endbibitem

\bibitem{embrechts2013modelling}
\begin{bbook}
\bauthor{\bsnm{Embrechts}, \binits{P.}},
\bauthor{\bsnm{Kl{\"u}ppelberg}, \binits{C.}},
\bauthor{\bsnm{Mikosch}, \binits{T.}}:
\bbtitle{Modelling Extremal Events: for Insurance and Finance}
vol. \bseriesno{33}.
\bpublisher{Springer},
\blocation{Germany}
(\byear{2013})
\end{bbook}
\endbibitem

\bibitem{abadie2008european}
\begin{barticle}
\bauthor{\bsnm{Abadie}, \binits{L.M.}},
\bauthor{\bsnm{Chamorro}, \binits{J.M.}}:
\batitle{{European CO2 prices and carbon capture investments}}.
\bjtitle{Energy Economics}
\bvolume{30}(\bissue{6}),
\bfpage{2992}--\blpage{3015}
(\byear{2008})
\end{barticle}
\endbibitem

\bibitem{boomsma2012renewable}
\begin{barticle}
\bauthor{\bsnm{Boomsma}, \binits{T.K.}},
\bauthor{\bsnm{Meade}, \binits{N.}},
\bauthor{\bsnm{Fleten}, \binits{S.-E.}}:
\batitle{Renewable energy investments under different support schemes: A real
  options approach}.
\bjtitle{European Journal of Operational Research}
\bvolume{220}(\bissue{1}),
\bfpage{225}--\blpage{237}
(\byear{2012})
\end{barticle}
\endbibitem

\bibitem{ernstsen2018valuation}
\begin{barticle}
\bauthor{\bsnm{Ernstsen}, \binits{R.R.}},
\bauthor{\bsnm{Boomsma}, \binits{T.K.}}:
\batitle{Valuation of power plants}.
\bjtitle{European Journal of Operational Research}
\bvolume{266}(\bissue{3}),
\bfpage{1153}--\blpage{1174}
(\byear{2018})
\end{barticle}
\endbibitem

\bibitem{aerts2013low}
\begin{barticle}
\bauthor{\bsnm{Aerts}, \binits{J.C.}},
\bauthor{\bsnm{Lin}, \binits{N.}},
\bauthor{\bsnm{Botzen}, \binits{W.}},
\bauthor{\bsnm{Emanuel}, \binits{K.}},
\bauthor{\bparticle{de} \bsnm{Moel}, \binits{H.}}:
\batitle{{Low-probability flood risk modeling for New York City}}.
\bjtitle{Risk Analysis}
\bvolume{33}(\bissue{5}),
\bfpage{772}--\blpage{788}
(\byear{2013})
\end{barticle}
\endbibitem

\bibitem{sharpe1964capital}
\begin{barticle}
\bauthor{\bsnm{Sharpe}, \binits{W.F.}}:
\batitle{Capital asset prices: A theory of market equilibrium under conditions
  of risk}.
\bjtitle{The Journal of Finance}
\bvolume{19}(\bissue{3}),
\bfpage{425}--\blpage{442}
(\byear{1964})
\end{barticle}
\endbibitem

\bibitem{lintner1964optimal}
\begin{barticle}
\bauthor{\bsnm{Lintner}, \binits{J.}}:
\batitle{Optimal dividends and corporate growth under uncertainty}.
\bjtitle{The Quarterly Journal of Economics}
\bvolume{78}(\bissue{1}),
\bfpage{49}--\blpage{95}
(\byear{1964})
\end{barticle}
\endbibitem

\bibitem{mossin1966equilibrium}
\begin{botherref}
\oauthor{\bsnm{Mossin}, \binits{J.}}:
Equilibrium in a capital asset market.
{Econometrica: Journal of the Econometric Society},
768--783
(1966)
\end{botherref}
\endbibitem

\bibitem{Dixit1994}
\begin{bbook}
\bauthor{\bsnm{Dixit}, \binits{A.K.}},
\bauthor{\bsnm{Pindyck}, \binits{R.S.}}:
\bbtitle{Investment Under Uncertainty}.
\bpublisher{Princeton University Press},
\blocation{Princeton, New Jersey}
(\byear{1994})
\end{bbook}
\endbibitem

\bibitem{codiga2011unified}
\begin{botherref}
\oauthor{\bsnm{Codiga}, \binits{D.L.}}:
Unified tidal analysis and prediction using the utide matlab functions
(2011)
\end{botherref}
\endbibitem

\bibitem{weinkle2018normalized}
\begin{barticle}
\bauthor{\bsnm{Weinkle}, \binits{J.}},
\bauthor{\bsnm{Landsea}, \binits{C.}},
\bauthor{\bsnm{Collins}, \binits{D.}},
\bauthor{\bsnm{Musulin}, \binits{R.}},
\bauthor{\bsnm{Crompton}, \binits{R.P.}},
\bauthor{\bsnm{Klotzbach}, \binits{P.J.}},
\bauthor{\bsnm{Pielke}, \binits{R.}}:
\batitle{{Normalized hurricane damage in the continental United States
  1900--2017}}.
\bjtitle{Nature Sustainability}
\bvolume{1}(\bissue{12}),
\bfpage{808}--\blpage{813}
(\byear{2018})
\end{barticle}
\endbibitem

\bibitem{newell2003discounting}
\begin{barticle}
\bauthor{\bsnm{Newell}, \binits{R.G.}},
\bauthor{\bsnm{Pizer}, \binits{W.A.}}:
\batitle{Discounting the distant future: how much do uncertain rates increase
  valuations?}
\bjtitle{Journal of Environmental Economics and Management}
\bvolume{46}(\bissue{1}),
\bfpage{52}--\blpage{71}
(\byear{2003})
\end{barticle}
\endbibitem

\bibitem{shao2015catastrophe}
\begin{barticle}
\bauthor{\bsnm{Shao}, \binits{J.}},
\bauthor{\bsnm{Pantelous}, \binits{A.}},
\bauthor{\bsnm{Papaioannou}, \binits{A.D.}}:
\batitle{Catastrophe risk bonds with applications to earthquakes}.
\bjtitle{European Actuarial Journal}
\bvolume{5}(\bissue{1}),
\bfpage{113}--\blpage{138}
(\byear{2015})
\end{barticle}
\endbibitem

\bibitem{johansson2014global}
\begin{barticle}
\bauthor{\bsnm{Johansson}, \binits{M.M.}},
\bauthor{\bsnm{Pellikka}, \binits{H.}},
\bauthor{\bsnm{Kahma}, \binits{K.K.}},
\bauthor{\bsnm{Ruosteenoja}, \binits{K.}}:
\batitle{{Global sea level rise scenarios adapted to the Finnish coast}}.
\bjtitle{Journal of Marine Systems}
\bvolume{129},
\bfpage{35}--\blpage{46}
(\byear{2014})
\end{barticle}
\endbibitem

\end{thebibliography}


\end{document}